\documentclass[aps,prmaterials,reprint,superscriptaddress, floatfix]{revtex4-2}

\usepackage{amsmath}
\usepackage{amssymb}
\usepackage{bm}
\usepackage{graphicx}
\usepackage{float} 
\usepackage{textcomp}
\usepackage{gensymb}
\usepackage[colorlinks=true, allcolors=blue]{hyperref}
\usepackage[utf8]{inputenc}

\newcommand{\kbf}{\mathbf{k}}

\newcommand{\xhat}{$\hat{\mathbf{x}}$}

\newcommand{\E}[1]{\textrm{E}[#1]}
\newcommand{\SNR}[1]{\textrm{SNR}[#1]}
\newcommand{\Var}[1]{\textrm{Var}[#1]}
\newcommand{\kc}{k_c}
\newcommand{\kx}{k_x}
\newcommand{\ky}{k_y}
\newcommand{\kz}{k_z}
\newcommand{\kr}{k_r}
\newcommand{\krho}{k_\rho}
\newcommand{\km}{k _{\textrm{max}}}
\newcommand{\Dt}{\Delta \theta}
\newcommand{\Dkt}{\Delta k_\theta}
\newcommand{\NPfig}{\vcenter{\hbox{$\mathord{\includegraphics[height=6ex]{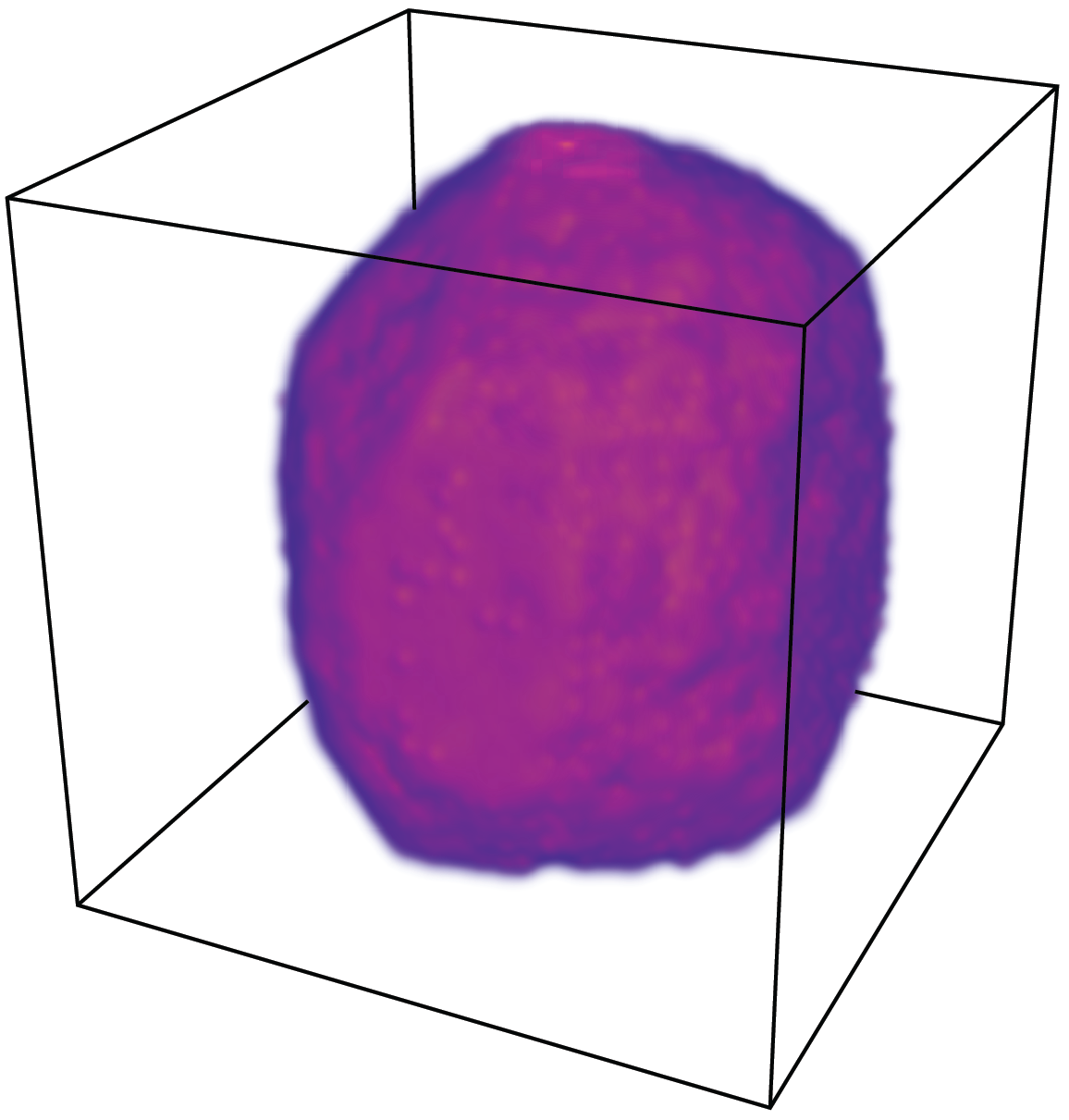}}$}}}
\newcommand{\PSFfig}{\vcenter{\hbox{$\mathord{\includegraphics[height=6ex]{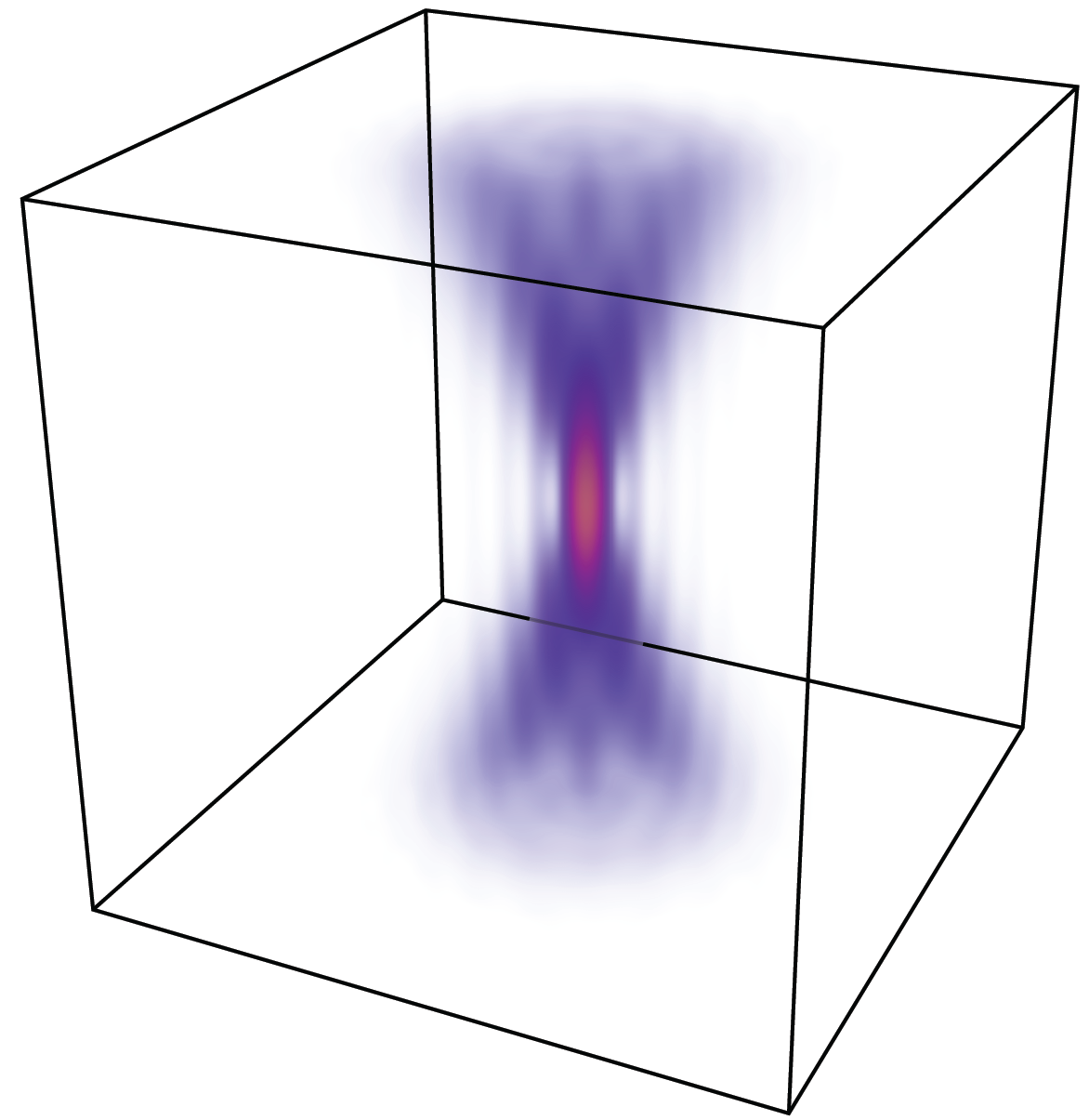}}$}}}
\newcommand{\planeFig}{\vcenter{\hbox{$\mathord{\includegraphics[height=6ex]{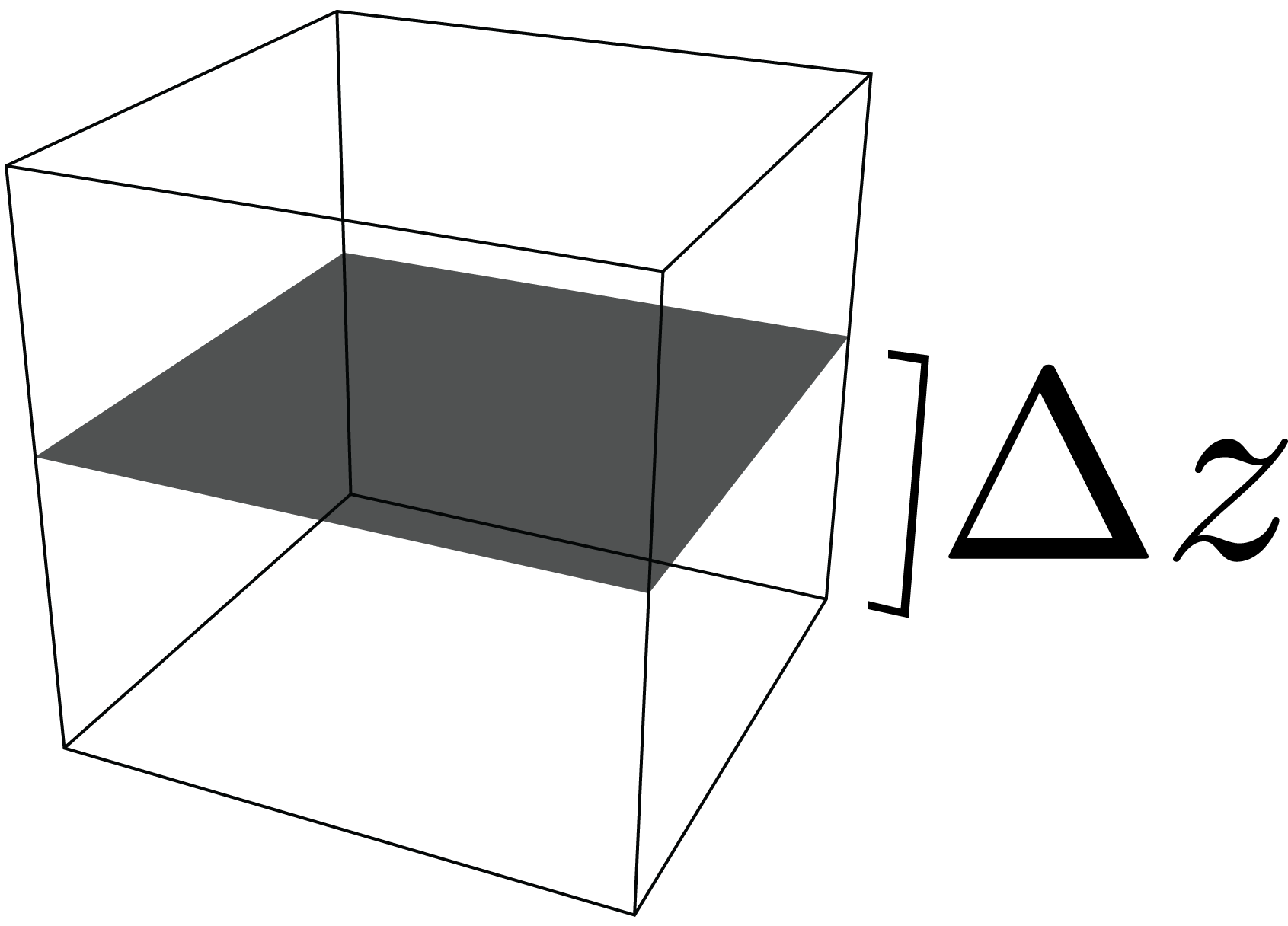}}$}}}
\newcommand{\NPkfig}{\vcenter{\hbox{$\mathord{\includegraphics[height=6ex]{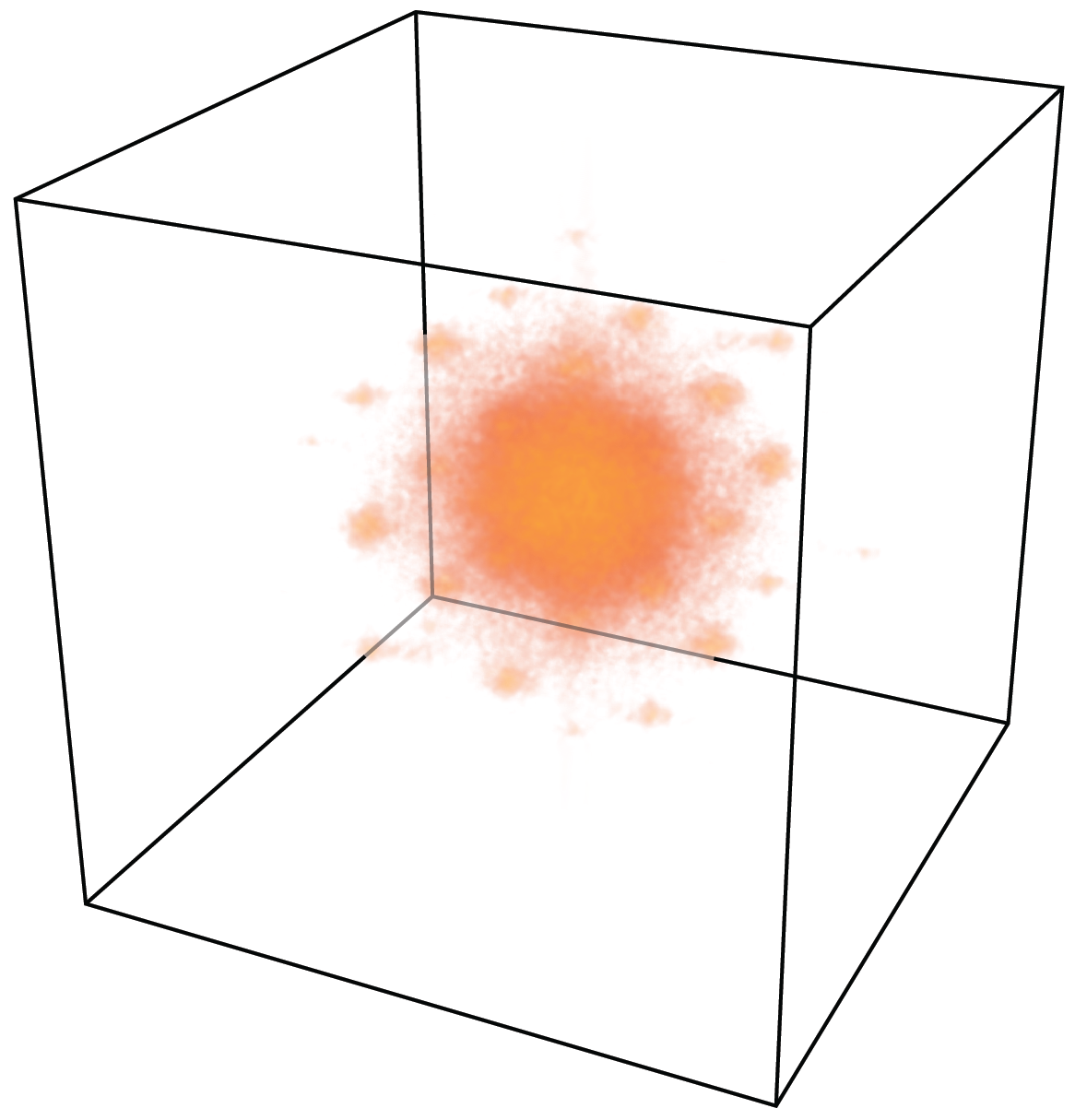}}$}}}
\newcommand{\CTFfig}{\vcenter{\hbox{$\mathord{\includegraphics[height=6ex]{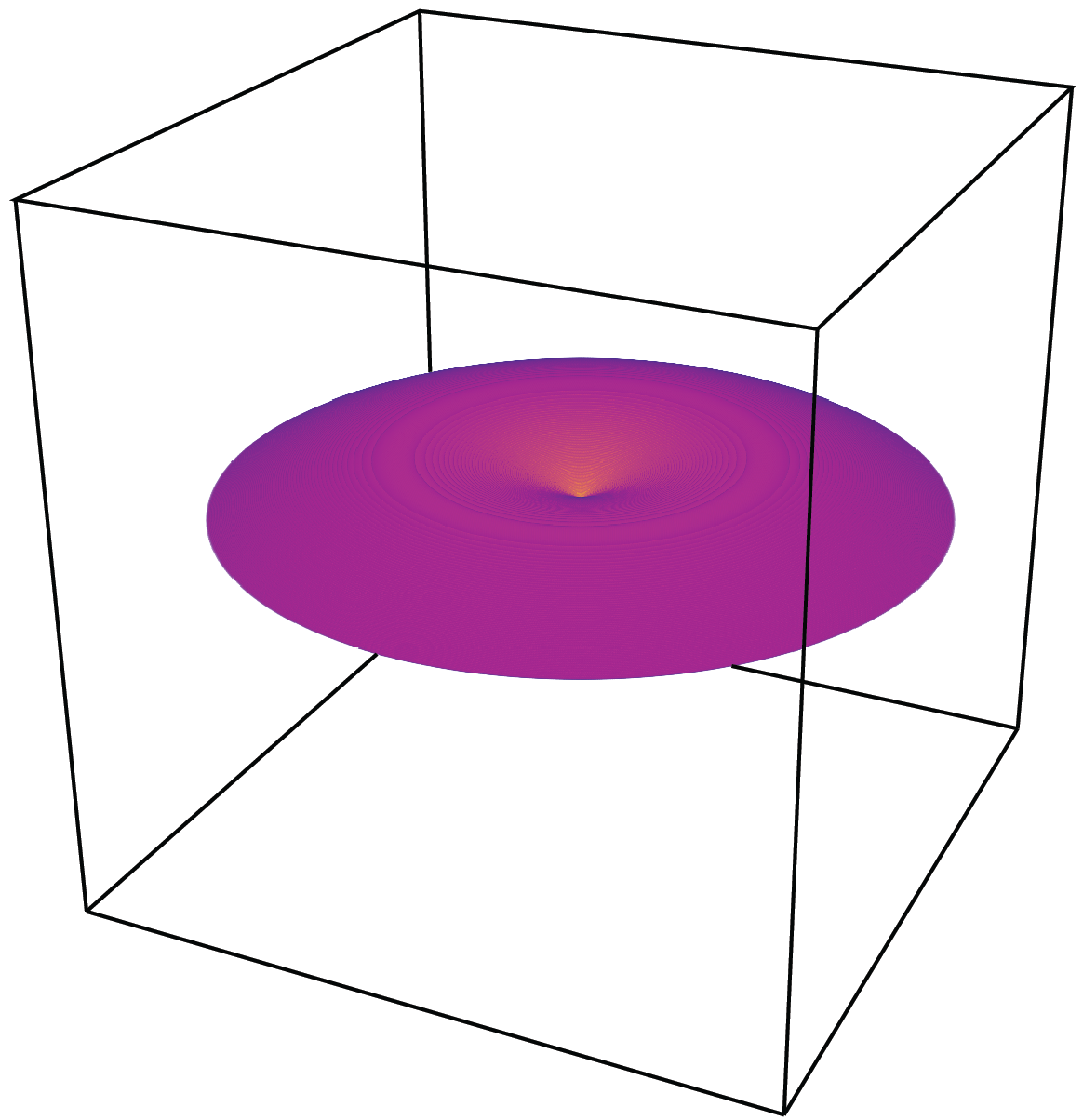}}$}}}
\newcommand{\rodFig}{\vcenter{\hbox{$\mathord{\includegraphics[height=6ex]{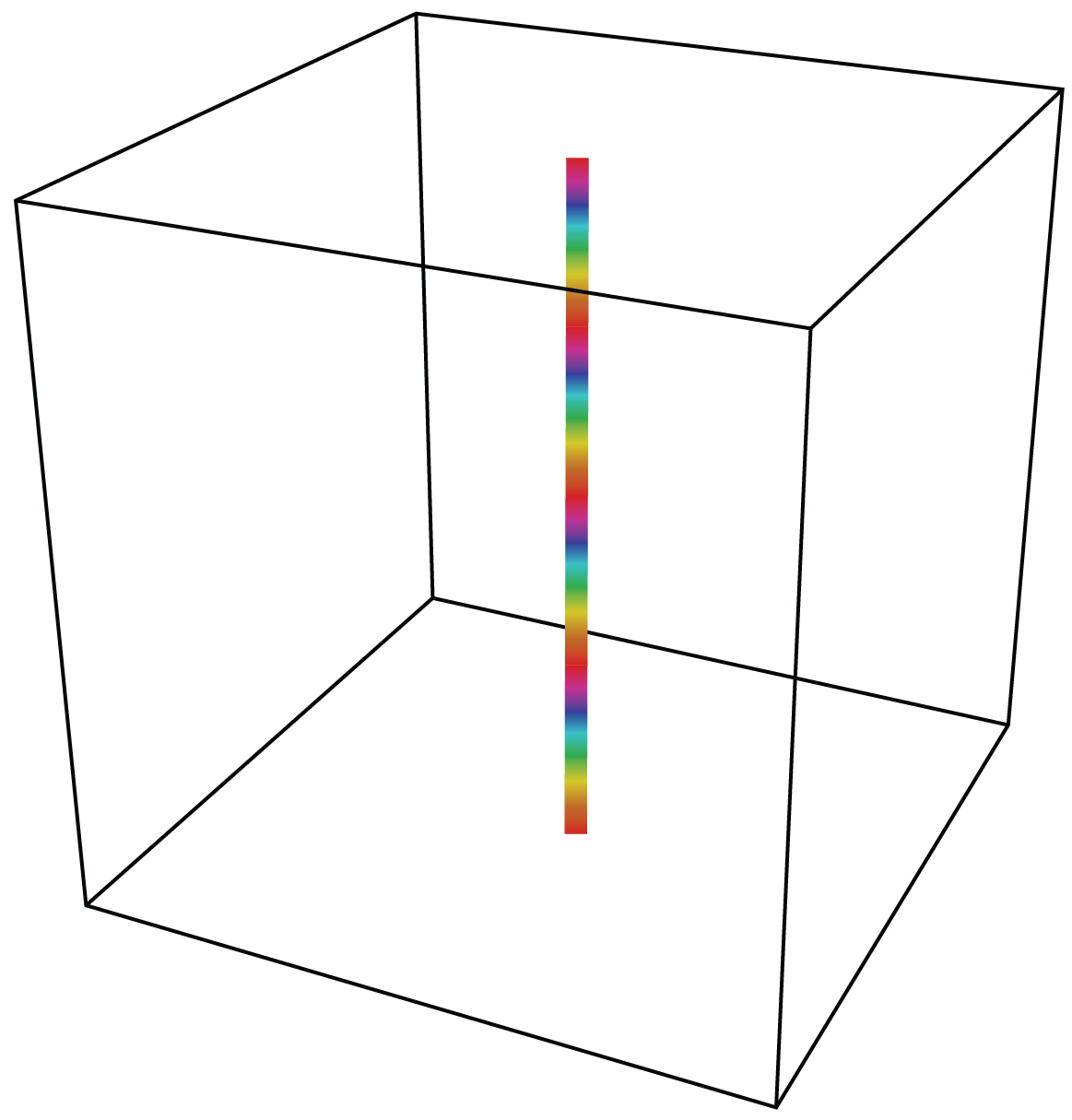}}$}}}

\DeclareRobustCommand{\rchi}{{\mathpalette\irchi\relax}}
\newcommand{\irchi}[2]{\raisebox{\depth}{$#1\chi$}} 

\begin{document}
\title{The Limits of Resolution and Dose for Aberration-Corrected Electron Tomography}
\author{Reed Yalisove}
\thanks{These two authors contributed equally}
\author{Suk Hyun Sung}
\thanks{These two authors contributed equally}
\affiliation{Department of Materials Science and Engineering, University of Michigan, Ann Arbor, Michigan 48109, USA}
\author{Peter Ercius}
\affiliation{The Molecular Foundry, Lawrence Berkeley National Laboratory, Berkeley, California 94720, USA}
\author{Robert Hovden}
\affiliation{Department of Materials Science and Engineering, University of Michigan, Ann Arbor, Michigan 48109, USA}
\affiliation{Applied Physics Program, University of Michigan, Ann Arbor, Michigan 48109, USA}
\email{hovden@umich.edu}
\date{\today}

\begin{abstract} 
Aberration-corrected electron microscopy can resolve the smallest atomic bond-lengths in nature. However, the high-convergence angles that enable spectacular resolution in 2D have unknown 3D resolution limits for all but the smallest objects ($< \sim$8 nm). We show aberration-corrected electron tomography offers new limits for 3D imaging by measuring several focal planes at each specimen tilt. We present a theoretical foundation for aberration-corrected electron tomography by establishing analytic descriptions for resolution, sampling, object size, and dose---with direct analogy to the Crowther-Klug criterion. Remarkably, aberration-corrected scanning transmission electron tomography can measure complete 3D specimen structure of unbounded object sizes up to a specified cutoff resolution. This breaks the established Crowther limit when tilt increments are twice the convergence angle or smaller. Unprecedented 3D resolution is achievable across large objects. Atomic 3D imaging (1 \AA) is allowed across extended objects larger than depth-of-focus (e.g. $>$ 20 nm) using available microscopes and modest specimen tilting ($<$ 3\degree). Furthermore, aberration-corrected tomography follows the rule of dose-fractionation where a specified total dose can be divided among tilts and defoci.
\\
\\
\textit{125 character summary: 
Unprecedented 3D resolution of large specimens established by novel analytic limits for aberration-corrected electron tomography.}
\end{abstract}


\maketitle
\section{Introduction}

\begin{figure*}
    \centering
    \includegraphics[width = 443pt]{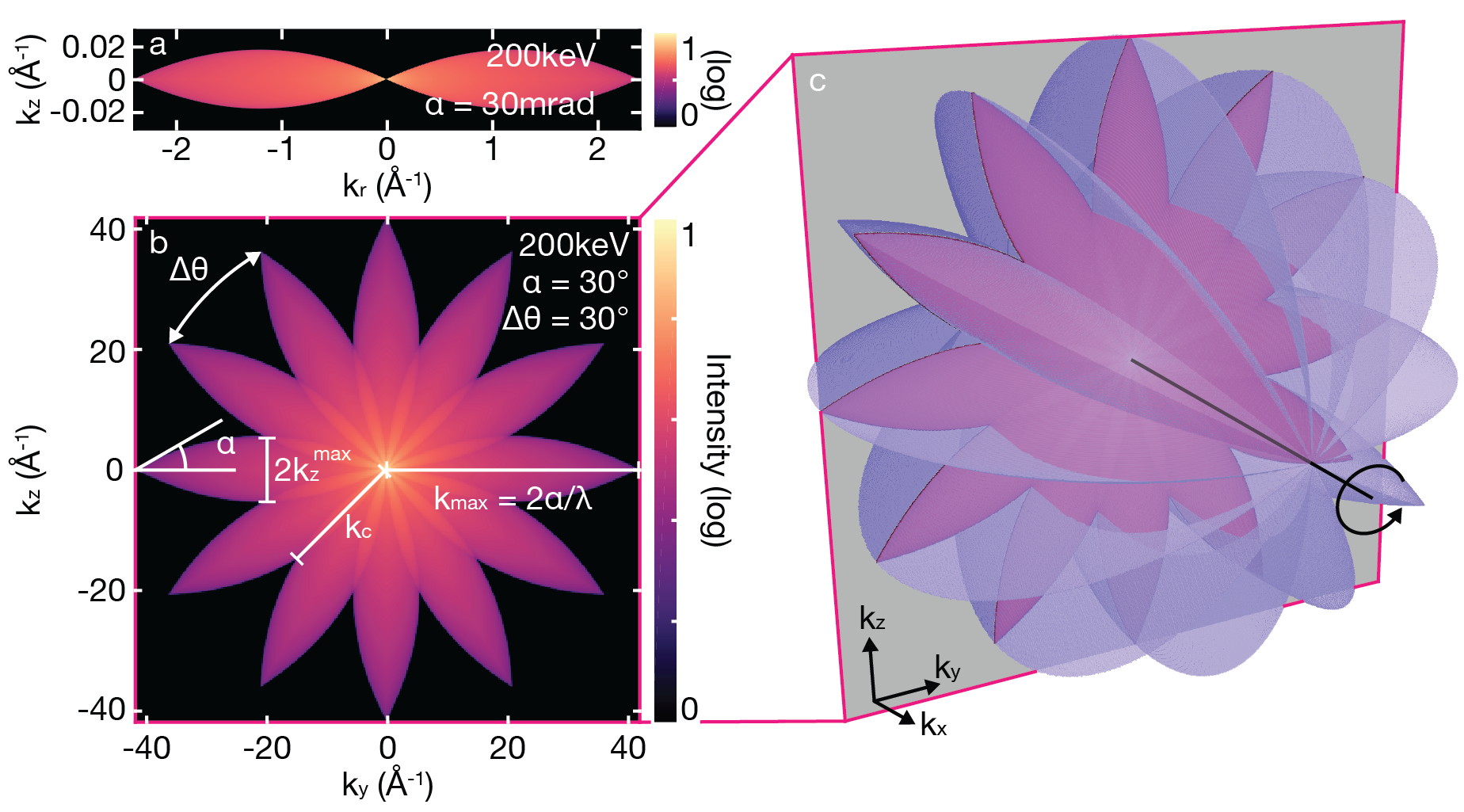}
    \caption{\textbf{3D contrast transfer function tomography using aberration-corrected scanning transmission electron microscopy.} a) Through-focal CTF for typical aberration-corrected STEM ($\alpha$ = 30mrad, 200keV). b) Internal structure of the tomographic CTF at $k_x = 0$ is highlighted. Each petal-shaped lobe represents a single through-focal CTF. Complete transfer of information is guaranteed within a spherical radius $k_c$. c) 3D CTF of through-focal tomography (tilt axis, \xhat). Blue shell denotes the information transfer limit in 3D. $\alpha$ and $\Delta \theta$ are exaggerated to 30\degree. The plane slices through the CTF at $k_x = 0$.}
    \label{fig:CTFshape}
\end{figure*}

Among the greatest goals in experimental science is to directly measure the complete 3D arrangement of atoms materials. However, fundamental sampling limits and an inextricable connection between lateral resolution and depth-of-focus have strictly prohibited 3D atomic measurement of extended materials. The theoretical limits of electron tomography have long been defined by (1) the Crowther-Klug criterion, which relates 3D resolution to the number of projections acquired~\cite{Klug_Crowther_1972}, and (2) the dose fractionation theorem~\cite{Hergerl_Hoppe_1976}, which states that the significance of a reconstructed object is independent of the distribution of dose. These limits were developed on the assumption that each image in a tomographic tilt series gives a perfect projection of the specimen. Over the last half century, this assumption has sufficed for microscopes where the depth-of-focus is large relative to the specimen\cite{Rosier_1968, Midgley_2001, Scott_2012, Yang_2017}. 

However, the defined limits of tomography are invalid for the new era of aberration-corrected scanning transmission electron microscopy (STEM) where highly-convergent electron beams confined to sub-\AA ngstrom lateral dimensions~\cite{Krivanek_1999} provide routine atomic imaging in 2D~\cite{Batson_2002,Nellist_2004,Muller_2008}. These revolutionary microscopes no longer provide simple projections of a specimen~\cite{Behan_2009,Hovden_2011, yang_imaging_2015} and tomography fails for objects larger than the depth-of-focus ($< \sim$8 nm). Hovden, et al. experimentally showed overcoming the limitations of aberration-corrected STEM tomography requires collecting a through-focal image stack at every specimen tilt~\cite{Hovden_Ercius_2014}. Although experimentally demonstrated, the theoretical limits of aberration-corrected tomography remain undefined. Understanding the tradeoff between resolution, object size, sampling, and dose using highly-convergent beams demands a new theoretical definition.

Here we present a theoretical foundation for aberration-corrected electron tomography that establishes analytic descriptions for resolution, sampling, and object size. We show that aberration-corrected tomography can far exceed the resolution limits of traditional tomography and breaks the conventional Crowther-Klug criteria.

The 3D contrast transfer function (CTF) for aberration-corrected STEM tomography distinctly measures a volume of information made from a superposition of toroids with petal-shaped cross-sections (Fig.~\ref{fig:CTFshape}) at every specimen tilt. A remarkable feature of the 3D CTF is the overlapped regions that permit complete 3D information collection up to a specified resolution ($1/k_c$)---unachievable with conventional tomography. This occurs when the incremental tilt angle ($\Delta \theta$) becomes smaller than twice the beam convergance semi-angle ($2\alpha$) which is typical to instrument operation ($\alpha>$ 25mrad) and sampling ($\Delta \theta<$ 2\degree). This complete information transfer breaks the Crowther-Klug relationships and the maximum reconstructable object size is unlimited up to a critical resolution ($1/k_c$). Beyond this critical resolution, Crowther-like tradeoffs define the maximum object size ($D$) allowed at a given 3D resolution ($d$). With more specimen tilts and higher convergence angles, 3D resolution improves quickly and the maximum object size increases dramatically. Using a tilt increment that matches the convergence angle ($\Delta \theta = \alpha$) any object size can be resolved with 3D resolution at $\sim$50\% of the microscope's diffraction limited resolution.

Despite the large amount of image data required by aberration-corrected electron tomography, the dose can be chosen to mitigate total specimen exposure. Extending the dose fractionation arguments presented by Hoppe\cite{Hergerl_Hoppe_1976} and Saxton\cite{saxberg_quantum_1981}, we show aberration-corrected electron tomography allows tunable dose allocation across any number of tilts and focal planes when oversampled. Thus, the signal-to-noise ratio of a 3D reconstruction is only dependent on the total dose imparted.

The relationships defined herein for aberration-corrected tomography supplant the Crowther-Klug criterion\cite{Crowther_Klug_1970, Klug_Crowther_1972} and the dose-fractionation theorem\cite{Hergerl_Hoppe_1976, saxberg_quantum_1981} that have long defined traditional tomographic techniques.

\section{Background}

In 1970, Crowther et al. established the fundamental tradeoff between 3D resolution, specimen size, and the number of projections measured\cite{Crowther_Klug_1970}. Bracewell and Riddle showed the same relationship for radio astronomy in 1967\cite{Bracewell_1967}. With evenly spaced specimen tilts about a single axis of rotation, the expression is compactly stated: $d = \pi D / N$, where $d$ is the smallest resolved feature size in three-dimensions, $D$ is object size, and $N$ is the number of projections acquired with equal angular spacing ($\Delta \theta = \pi/N$). This sampling criterion is the most stringent requirement that can be adopted and ensures specimen features are measured and the entire reconstruction is equally sharp and free of aliasing.

Conceptually, 3D resolution is limited by tomography's inability to collect complete information about the specimen. Projection images at each tilt map to a plane of information in frequency space (k-space)---as defined by the projection slice theorem\cite{Bracewell_1990}. The missing information between planes limits the 3D resolution and object size of a tomographic reconstruction. For specimens tilted about a single axis of rotation, the planes of information intersect along one axis ($k_x$) on a cylindrical coordinate system as illustrated in Supplemental Figure~\ref{fig:convCTF3D}. Sampling is maximal radially and along the axis of rotation, however undersampling occurs azimuthally along $k_{\theta}$ and worsens at higher frequencies ($k_r$). The azimuthal gap between adjacent measurement planes ($\Delta k_{\theta} \approx k_r \cdot \Delta\theta$) limits the largest resolvable object: $\Delta k_{\theta} = 1/D$. Thus, collecting more specimen tilts reduces the distance between measurement planes and allows higher resolution (larger $k_r$) or larger object sizes in 3D. This theorem is well suited for traditional S/TEM tomography where the depth-of-focus is larger than the object size (Supplemental Figure~\ref{fig:psfctf}).

Six years after Crowther et al. defined the resolution and sampling limits of tomography, Hegerl and Hoppe established a dose fractionation property for tomography, which Saxberg and Saxton further refined after debate \cite{Hergerl_Hoppe_1976, saxberg_quantum_1981}. Their work showed when an object is sufficiently sampled (i.e. better than Crowther-Klug requirements) the signal-to-noise ratio (SNR) of a reconstruction depends only on the total dose imparted. Maintaining an equivalent total dose, one may divide that dose across more or fewer projections. Both derivations are based on weak-contrast imaging---however we will show a weak-contrast approximation is not required.

\section{Contrast Transfer Function of Aberration-Corrected Tomography}

\begin{figure*}
    \centering\includegraphics[width = 509pt]{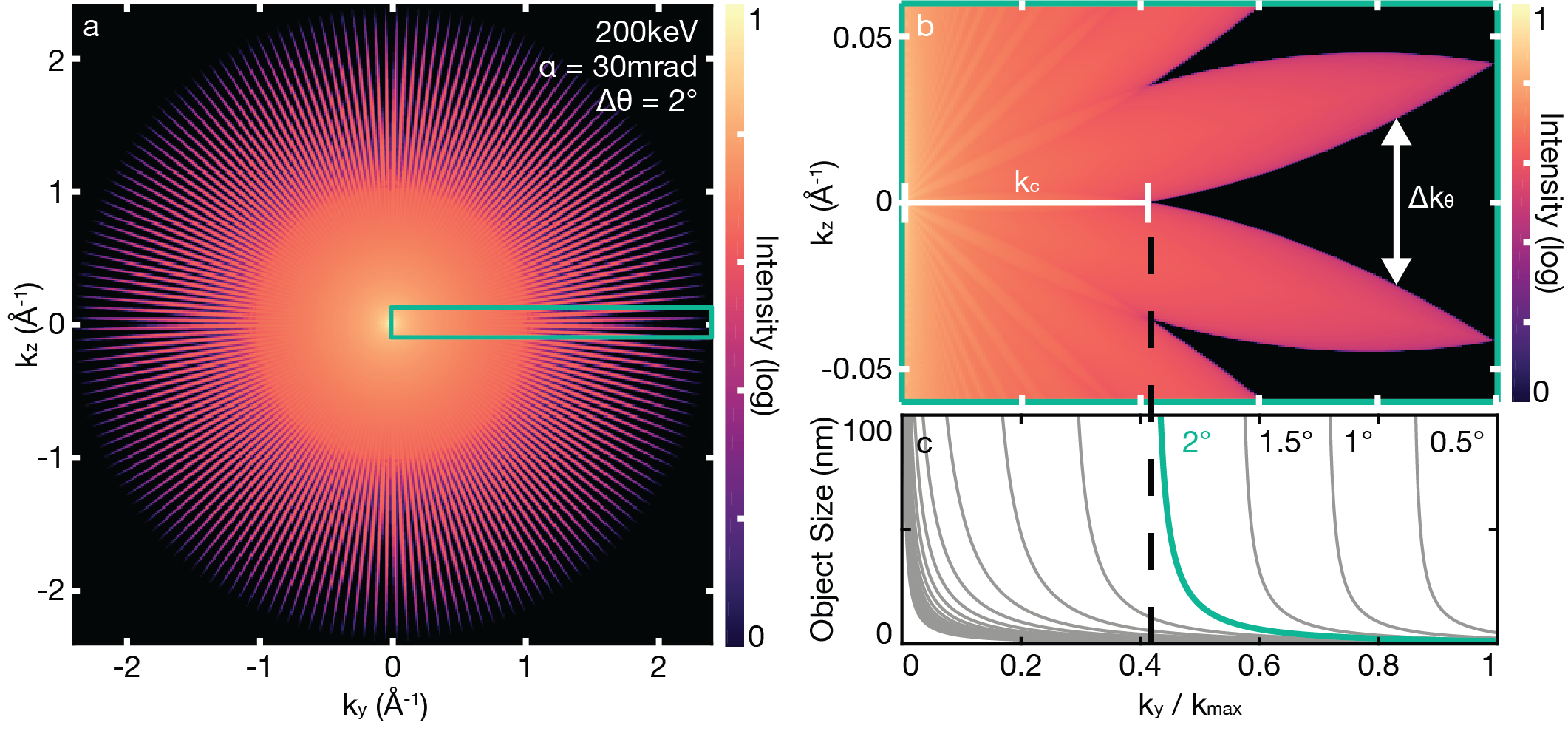}
    \caption{\textbf{Relationships between tilt angle, resolution, and maximum object size} The CTF of aberration-corrected tomography hosts overlap regions which permit complete information transfer—and therefore unlimited object size with resolution $d = 1/\kc$. a) Full tomographic CTF in cross-section for 200 keV, 30 mrad, 2\degree{} tilts. b) Subregion of full tomographic CTF highlights the maximum frequency of complete information transfer, $\kc$, and $\Dkt$ denotes separation between through-focal CTFs at each tilt. c) Spatial frequency vs. maximum object size for several tilt step-sizes with teal curve matching conditions in a). For $\ky < \kc$, the maximum object size is unbounded.}
    \label{fig:NewCrowther}
\end{figure*}

Due to the highly convergent nature of aberration-corrected electron beams, Crowther's derivation of sampling requirements do not hold for aberration-corrected tomography. The CTF is no longer a 2D plane; its form becomes a 3D toroid (Fig.~\ref{fig:CTFshape}a) derived analytically by Intaraprasonk, Xin, Muller \cite{Varat_Muller_2008} and reproduced in Supplemental Information~\ref{CTFderive}. Thus for aberration-corrected tomography, the projection planes in k-space are replaced by toroidal CTFs. Figure~\ref{fig:CTFshape}b,c show the total tomographic CTF in 3D. Summing a rotated set of toroidal CTFs describes the region of k-space from which information is collected in aberration-corrected tomography:\par
\begin{equation}
    H(\kr, \kz) = \sum _\theta h (\kr^\theta, \kz^\theta) \label{ctfSum}
\end{equation}
Here $h(\kr^\theta, \kz^\theta)$ is the radially symmetric toroidal CTF from a single specimen tilt, $\theta$, measured by through-focal imaging. For an aberration-free beam this through-focal CTF is:
\begin{gather}
    h (\kr^\theta, \kz^\theta) = \frac{1}{2 \pi^3 \alpha \kr^\theta} \sqrt{1 - \left ( \frac{\kr^\theta \lambda}{2 \alpha} + \frac{|\kz^\theta|}{\alpha \kr^\theta} \right ) ^2 } \label{ctfeq}\\
    |\kz^\theta| \leq \frac{\lambda}{2} \kr^\theta \left (\frac{2 \alpha}{ \lambda} - \kr^\theta \right ) \label{ctfbound}
\end{gather}
 where $\alpha$ is the convergence semi-angle of the electron beam, $\kr$ is the radial frequency, and $\lambda$ is the wavelength of the electron. By parameterizing the equation for the through-focal CTF (Eq.~\ref{ctfeq}), we express a CTF that has been tilted about an axis ($\kx$) perpendicular to the optical axis ($\kz$) in k-space. This rotation gives $\kr^\theta = \kr \cos(\theta) - \kz \sin(\theta)$ and $\kz^\theta =\kr \sin(\theta) + \kz \cos(\theta)$ for $\theta \in (-\pi/2, \pi/2]$. Each CTF is bounded by Equation~\ref{ctfbound}. These equations describe an aberration-free probe formed by an objective aperture that limits aberrations and sets the convergence semi-angle (Supplemental Information~\ref{CTFderive}).
 
The CTF for aberration-corrected tomography (Eq.~\ref{ctfSum}--\ref{ctfbound}) is shown in Figure~\ref{fig:NewCrowther}a,b for a 200keV electron beam with 30mrad semi-convergence angle and a $2\degree$ tilt interval about a single axis of rotation.

\section{3D Resolution and Object Size Limits}

For aberration-corrected tomography, the sampling requirements that limit object size and resolution are determined by the missing information between the toroidal bounds of the through-focal CTF (Eq.~\ref{ctfbound}). For infinitely large objects, measuring specimen structure at a single frequency ($\mathbf{k'}$) requires the information to lie within the tomography CTF. However, for objects of finite size, $D$, the condition loosens and CTF bounds only need to measure the neighborhood, $\mathbf{k'}+\bm{\delta}$ where $|2\bm{\delta}| = 1/D$ (Appendix Fig. \ref{fig:CTFdimDiagram}). Thus, the size of missing information in k-space limits the detectable object size. Similar to the analysis by Crowther and Klug, we aim to calculate the maximum object size for a tomographic reconstruction by calculating the k-space distance, $\Delta k_{\theta}$, between the information collected at sequential specimen tilts. The non-planar geometry of aberration-corrected tomography reduces distances of missing information in k-space that permit measurement of larger real space objects at higher resolutions than the Crowther-Klug limit.

The strictest resolution requirement ensures measurement of objects of any shape or symmetry without any prior information. Although the tomographic CTF is non-isotropic and resolution is higher along the axis of rotation, we define 3D resolution by the worst measurable resolution. For single tilt-axis tomography, undersampling occurs along $k_{\theta}$ and defines the largest k-space distance between adjacent through-focal CTFs. The missing information ($\Delta k_{\theta}$) increases at higher frequencies, $k_r$, and thus limits resolution ($d = 1/k_{r}$). We show (Appendix~\ref{ResolutionDerive}) the distance between measured information is maximal on the $\ky\kz$-plane (Fig.~\ref{fig:CTFshape}) and is used to calculate the limits for aberration-corrected tomography.

The most striking feature of the aberration-corrected tomography CTF is the continuum of information it can permit. From Figure~\ref{fig:CTFshape}b we see that with small tilt increments and large convergence angles, adjacent through-focal CTFs will overlap, allowing complete information transfer up to a critical frequency, $\kc$:
\begin{equation}
    \kc = \frac{2 \alpha - \Dt}{\lambda}
    \label{eq:kc}
\end{equation}
 where $\Delta\theta$ is the angular spacing between specimen tilts. Equation~\ref{eq:kc} defines this critical frequency under a first-order small-angle approximation as derived in Appendix~\ref{ResolutionDerive} and is valid when $k_c$ is positive (i.e. $\Dt < 2 \alpha$). 
 
 This critical frequency splits the problem into two regimes, so the resolution limit on object size is defined piece-wise. For $\kr \leq \kc$, the structure of the specimen is completely measured---this corresponds to unbounded maximum object sizes. For $\kr > \kc$, there is a finite distance between adjacent regions of information (Appendix Eq.~\ref{eq:CTFdistanceNoApprox}), which relates the maximum frequency, $k_y$, in a reconstruction to the maximum object size, $D$ (shown in Figure~\ref{fig:NewCrowther}c). The piece-wise expression is:\par
\begin{equation}
    \label{eq:fov_real}
    D = \left \{\begin{array}{ll}
        \frac{d^2}{\lambda (1 - k_c d)}, & \frac{\lambda}{2} < d < \frac{1}{k_c}\\
        \infty, & \frac{1}{k_c} \leq d < \infty
        \end{array}
        \right .
\end{equation}
Equation~\ref{eq:fov_real} defines a new limit relating resolution, object size, and sampling for aberration-corrected electron tomography, analogous to the Crowther-Klug limit for conventional tomography. It shows higher beam energies (i.e. smaller wavelengths) and higher convergence angles allow higher resolution and larger object sizes.

\begin{figure}
    \centering
    \includegraphics[width = 231pt]{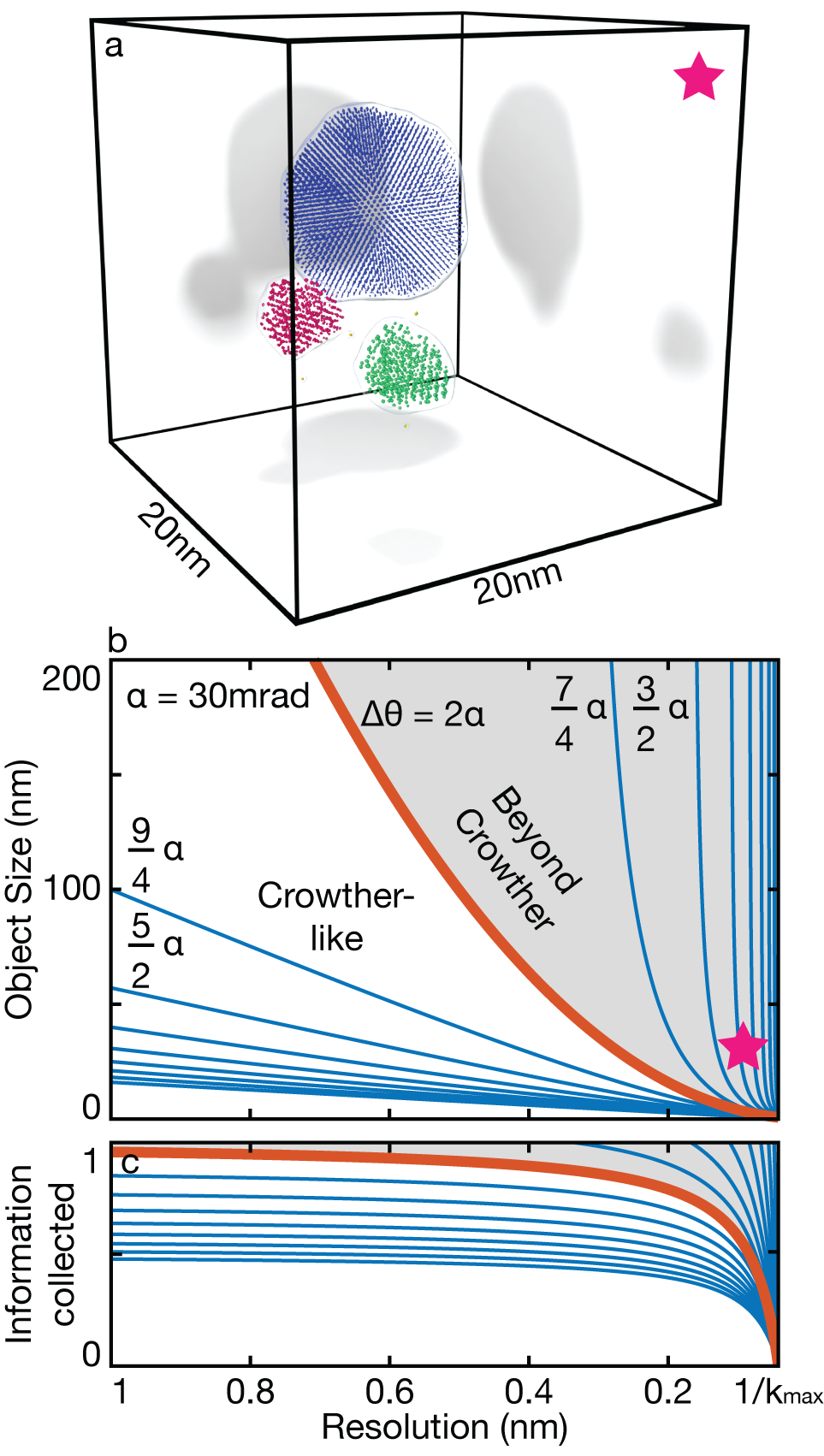}
    \caption{\textbf{Aberration-corrected electron tomography enables unprecedented high-resolution of extended objects.} a) 3D atomic resolution tomography of three nanoparticles in a 20nm volume---reconstructed here from quantum mechanical scattering simulations. b) Tradeoff between 3D resolution and object size is plotted for different specimen tilt increments, $\Delta \theta$. The Crowther limit is surpassed when $\Delta \theta \leq 2\alpha$ (grey) such that objects of any size may be reconstructed. Star denotes the size and resolution of reconstruction in (a). c) The percent of information collected at each resolution.}
    \label{fig:resVobj}
\end{figure}

 Remarkably, when $\Delta \theta < 2 \alpha$, there is always a resolution at which an infinite object size can be reconstructed (neglecting multiple scattering). This behavior is not predicted by the Crowther criterion and exceeds the previously expected limits. As illustrated in Figure~\ref{fig:resVobj}, with even smaller tilt increments the reconstructable object size diverges at high resolution. Notably, when the tilt increment matches the convergence angle ($\Delta \theta = \alpha$) any object size can be resolved in 3D at $\sim$50\% of the microscope's diffraction limit. For $\Delta \theta = \alpha/2$ a 3D resolution at 75\% the diffraction limit is achievable for an unlimited object size. 

At higher resolution beyond the critical frequency, $k_c$, aberration-corrected tomography still outperforms traditional tomography due to the reduced missing information between lobes in the tomography CTF. Figure~\ref{fig:resVobj}b, shows the trade-off between object size and resolution is favorably non-linear. It provides the object size that can be reconstructed at a given resolution for different specimen tilt increments. For example, a 75 nm specimen, imaged with a 30 mrad convergence angle and 3\degree~(50 mrad) increments between tilts allows 2 \AA~resolution in 3D at 200 keV.

Moreover, atomic resolution imaging, with 1.5 \AA~resolution in 3D, is possible over a 15 nm object if sampled at 3\degree~using a 200 keV beam and 30 mrad convergence semi-angle. 3D atomic resolution imaging of extended objects has been computationally verified (Fig.~\ref{fig:resVobj}a) using quantum mechanical multiple scattering simulations of aberration-corrected tomography performed on crystalline nanoparticles within a 20 nm volume (See Supplemental Information~\ref{SuppMethods}). This simulation computed over 500 million elastically scattered electron wavefunctions to generate images at 13 defocus positions at each of 105 tilts, using over 15,000 GPU core hours. 

Despite some missing information at higher frequencies ($k_r>k_c$), a significant portion of k-space is measured and aberration-corrected tomography provides a superior reconstruction compared to traditional tomography. However, due to the finite periodic sampling of specimen tilts, aberration-corrected tomography may still permit weak aliasing. Aliasing can occur azimuthally at high radial frequencies, $k_r>k_c$, when reconstructed object sizes exceed the Crowther-Klug relation, even if the requirements for resolution (Eq.~\ref{eq:fov_real}) are met. Fortunately, if present, aliasing is substantially attenuated by the amount of information collected with aberration-corrected tomography. The intensity of azimuthal aliasing is proportional to the percentage of information collected at each radial frequency $k_r$ (plotted as a function of resolution in Figure~\ref{fig:resVobj}c). With more measured information, the reconstruction quality improves and aliasing becomes negligible. Thus, no aliasing occurs at low frequencies ($k_r<k_c$) and at the microscope's transfer limit ($k_r = k_{max}$) the aliasing is significant and matches traditional tomography.

\section{Defocus Sampling Requirement}
\label{sec:defocusSampling}
 At each specimen tilt, aberration-corrected electron tomography acquires a through-focal stack of images. This overcomes the limited depth-of-focus ($<\sim$8nm) where a single image cannot sufficiently measure large objects. With through-focal imaging, the focal plane incrementally moves through the entire object and measures specimen information within a through-focal CTF (Eq.~\ref{ctfeq}). The defocus step size, $\Delta z $, becomes an an additional sampling requirement.

The defocus step must be smaller than the microscope's depth-of-focus. This sampling requirement is described by the widest portion of a through-focal CTF ($k_z^\textrm{max}$) along the beam direction ($k_z$). The largest defocus step size is:

\begin{equation}
    \Delta z^{\textrm{max}} = \frac{1}{2 k_z^{\textrm{max}}} = \frac{\lambda}{\alpha^2}
    \label{eq:zNyquist}
\end{equation}
This equation is the well-known depth-of-focus relationship, and is analytically derived via wave optics in Appendix~\ref{app:defocusSample}.

Ideally, the focal range should not exceed the object being measured as images captured beyond the object bounds increases dose to the sample without adding information. The most dose-efficient measurement has a field-of-view and defocus range that matches the object size.

\section{Dose Fractionation}

Surprisingly, aberration-corrected tomography is not necessarily dose intensive. Expanding the dose fractionation theorem the total dose can be chosen and distributed among both specimen tilts and defoci. Hegerl and Hoppe's original construction for conventional electron tomography states that the SNR of reconstructed voxels depends only on the total dose imparted, not the distribution of dose. It assumes weak contrast imaging with additive noise~\cite{Hergerl_Hoppe_1976,saxberg_quantum_1981}. We use a more complete noise model with Poisson statistics~\cite{foi_2008}. For a Poisson limited signal, each noisy image $\widetilde{p}(x,y)$ of projected object $p(x,y)$ has a signal-to-noise ratio of $\SNR{\widetilde{p}(x,y)} = 1+ \rchi t \, p(x,y)$ for acquisition time $t$ and dose-rate $\rchi$ (See Appendix~\ref{app:noisemodel}). For tomography, the signal and noise variance from projections at each tilt add linearly to the final reconstruction because the noise from each image is uncorrelated. It shows the SNR for projection images---and tomographic reconstructions thereof---depends on both the dose and the specimen.

3D reconstruction quality is independent of the number of specimen tilts only when k-space is sufficiently sampled (i.e. oversampled) and the total dose is evenly distributed across equally spaced tilts. The same is true for aberration corrected tomography where the signal and noise from volumetric through-focal CTFs add linearly in k-space. The SNR of the final reconstruction depends on the total dose imparted onto the specimen not the number of specimen tilts---so long as k-space is sufficiently sampled. Notably, oversampling is guaranteed below $k_c$ (Eq.~\ref{eq:kc}). 

\begin{figure}[ht]
    \centering
    \includegraphics[width = 235pt]{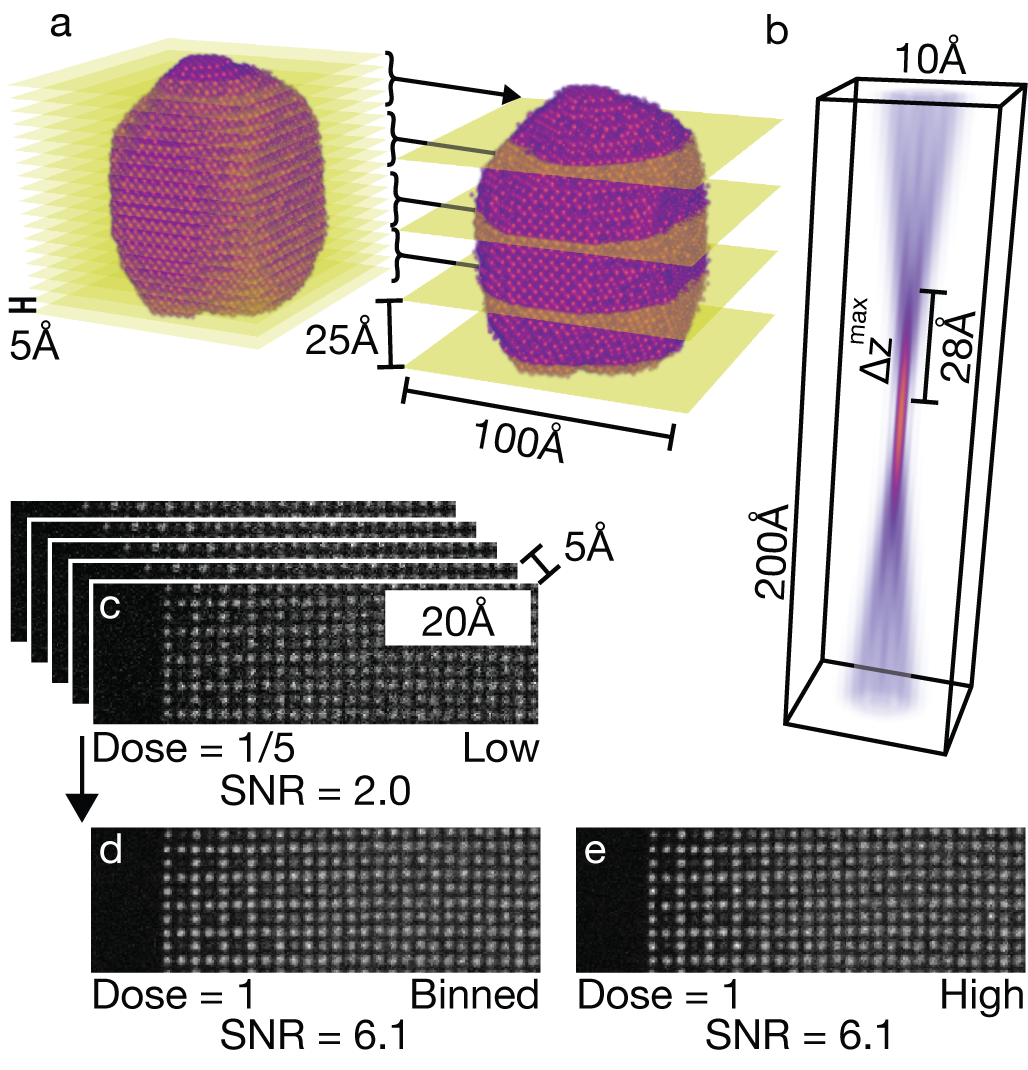}
    \caption{\textbf{Dose fractionation for through-focal acquisition} a) Oversampling of focal planes contains the same information as fewer focal planes, so long as the defocus sampling requirement is met. b) Defocus sampling requirement is set by the depth-of-focus ($\Delta z^{\textrm{max}}$, 28\AA{} for 200keV, 30mrad). Multislice simulation with Poisson noise shows the equivalence of SNR in c,d) the sum of adjacent low-dose defocused images and e) a single high dose image.}
    \label{fig:doseFrac}
\end{figure}
However, for aberration-corrected tomography, dose is not only divided among tilts, but also among defoci. Here, dose fractionation also holds for through-focal image acquisitions. SNR of a through-focal stack describes the quality of data at a given dose and dose distribution (See Appendix~\ref{app:doseFrac}). In an oversampled through-focal image stack ($\Delta z \ll \Delta z^{\textrm{max}}$), adjacent defocused images can be summed without loss of information (Fig.~\ref{fig:doseFrac}a,b). The SNR after summing $M$ adjacent images ($\SNR{\sum_{\Delta z}\widetilde{p}(x,y,z_f+\Delta z)} = 1 + M \rchi t \, \bar{p}(x,y)$) across defoci $z_f + \Delta z$ with dose-per-image $\rchi t$ matches that of a single image, $\bar{p}(x,y,z_f)$, taken with the same total dose ($M\rchi t$). It is the total dose across all images, not the number of images, that determines the SNR of useful information so long as the defocus sampling requirement is met (Eq.~\ref{eq:zNyquist}). 

Fully quantum-mechanical multislice simulations with Poisson noise demonstrate the dose fractionation theorem for through-focal imaging in Figure~\ref{fig:doseFrac}. A simulated high-dose image (Fig.~\ref{fig:doseFrac}e) and a binned through-focal stack with same total dose (Fig.~\ref{fig:doseFrac}c,d) have comparable SNR and carry the same information. We expand this over the full range of defoci. When the through-focal stack is evenly oversampled along defocus, the reconstruction quality is dependent only on the total dose, not the distribution.

Thus, aberration-corrected tomography does not inherently require high doses. The desired total dose for a given specimen determines the SNR of a 3D reconstruction. The total dose may be chosen and divided across tilts and defoci, so long as all sampling requirements are sufficiently met. The traditional dose fractionation theorem is upheld in aberration-corrected tomography but has the added dimension of defocus sampling. Unfortunately we anticipate aberration corrected tomography will still adhere to traditional dose requirements where 3D resolution scales inversely with dose$^{1/4}$ \cite{saxberg_quantum_1981, Mcewen_2002} and atomic resolution requires substantial beam exposure.

\section{Conclusion}

Aberration-corrected tomography's volumetric CTF breaks the traditional sampling requirements for object size and resolution as famously set by Crowther and Klug~\cite{Klug_Crowther_1972}. Accounting for the highly-convergent imaging probes, a novel limit on resolution, object size, and sampling is presented in Equation~\ref{eq:kc} and~\ref{eq:fov_real}. Up to a critical spatial frequency, aberration-corrected tomography can reconstruct an object of any size, and above that frequency the limits on object size still exceed conventional tomography. This is critically significant for the next generation of electron microscopes with ever increasing convergence angles ($>$ 60 mrad) and diminishing depth-of-focus. Lastly, the signal-to-noise of a tomographic reconstruction is determined by the total dose of the measurement and that dose may be distributed among defoci and specimen tilt. 

Moreover, this work extends beyond scanning transmission electron tomography and is applicable to any incoherent linear imaging technique that uses highly convergent beams where the depth-of-focus is small compared to the 3D object size. 

With the theoretical limits defined herein, we can proceed to higher resolution across larger fields-of-view to know the atomic structure of extended specimens in all three dimensions.

\begin{acknowledgments}
We thank Yi Jiang for helpful discussions. R.Y., S.H.S. acknowledge support from the DOE BES(j) (Subaward No. K002192-00-S01). R.H. acknowledges support from the Keck Foundation. Simulations made use of the Advanced Research Computing Technology Services’ shared high-performance computing at the University of Michigan and the Molecular Foundry (supported by the Office of Science, U.S. Department of Energy under Contract No. DE-AC02-05CH11231). 
\end{acknowledgments}

\pagebreak
\clearpage
\bibliographystyle{apsrev4-2}
\bibliography{refs}

\clearpage
\appendix
\renewcommand\thefigure{\Alph{section}\arabic{figure}}
\setcounter{figure}{0}

\section{Resolution, Sampling, Object-Size Relationship for Aberration-Corrected Tomography}
\label{ResolutionDerive}
\begin{figure}[b]
    \centering
    \includegraphics[width = 232pt]{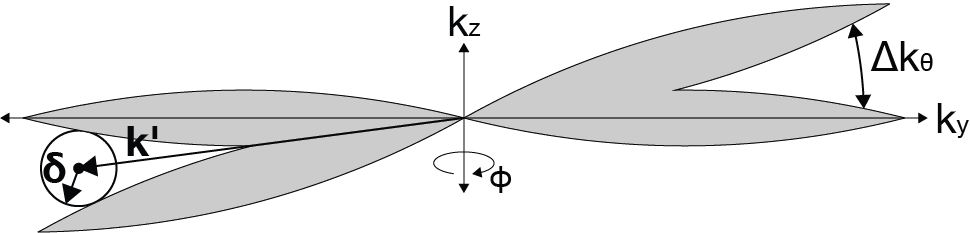}
    \caption{Illustration of two adjacent tilted CTFs in the plane of maximum separation. On the right, $\Dkt$ separates CTFs, and on the left, the neighborhood about $\mathbf{k'}$ with diameter $|2\bm{\delta}| = 1/D$ limits the maximum object size}
    \label{fig:CTFdimDiagram}
\end{figure}
The information measured in reciprocal space by aberration-corrected STEM tomography is described by a superposition of contrast transfer functions (CTFs) from each specimen tilt about a single axis of rotation (Eq.~\ref{ctfSum}-~\ref{ctfbound}). The maximum object size that can be reconstructed with a given resolution is determined by the arc length between adjacent tilted CTFs, $\Delta k_\theta$. The upper bound of a CTF in cylindrical coordinates about the $\kz$-axis ($\kr, \phi, \kz$) is
\begin{equation}
\label{eq:CTFbound}
    \kz = \frac{\lambda}{2} \kr \left (\frac{2 \alpha}{\lambda} - \kr \right )
\end{equation}
This upper bound is radially symmetric about the $\kz$-axis and the lower bound is a reflection across the $\kz=0$ plane. Equivalent points (same $\kx$ and $\ky$) on the bounds of a CTF in Cartesian coordinates are located at $\kbf_1= (k_x,k_y,k_z)$ and $\kbf_2= (k_x,k_y,-k_z)$.

Fixing $\kbf_1$ and rotating $\kbf_2$ by an angle $\Dt$ about the $\kx$-axis we find the vector between adjacent CTFs, $\Delta\bm{k}_\theta$.
\begin{align*}
    \Delta\bm{k}_\theta = \begin{pmatrix} 
    0 \\
    \ky (\cos \Dt - 1) + \kz \sin \Dt \\
    \ky \sin \Dt - \kz (\cos \Dt + 1)
    \end{pmatrix}
\end{align*}
The magnitude of this vector accurately approximates the arc length between equivalent points on two adjacent CTFs.
\begin{align*}
    \Dkt &= \left ( [ \ky ( \cos \Dt - 1 ) +  \kz \sin \Dt ]^ 2 +  \right . \\
    & \phantom{=} \left .  [ \ky \sin \Dt - \kz ( \cos \Dt + 1)]^2 \right )^{\frac{1}{2}}\\ 
    &= \sqrt{2} \left ( \ky ^2 (1 - \cos \Dt) - \right . \\
    &\phantom{=} \left . 2 \ky \kz \sin \Dt + \kz ^2 ( 1 + \cos \Dt) \right ) ^ \frac{1}{2}\\ 
    &= \sqrt{2} \left ( \ky \sqrt{1 - \cos \Dt} - \kz \sqrt{ 1 + \cos \Dt} \right )
\end{align*}
Transforming to cylindrical coordinates ($\ky = \kr \sin \phi$), substituting Equation~\ref{eq:CTFbound}, and using sine and cosine half-angle identities, the distance between two adjacent CTFs is
\begin{align*}
    \Dkt &= 2 \left [ \kr \sin \phi \sin \frac{\Dt}{2} + \frac{\lambda}{2} \kr \left (\kr - \frac{2 \alpha}{\lambda} \right ) \cos \frac{\Dt}{2} \right ]
\end{align*}
 Information is maximally sampled along $\kr$, $k_x$ and undersampled along $k_\theta$ due to a finite number of specimen tilts.  The strictest limit for resolution and object size is set by the path in k-space along which $\Dkt$ is largest.  We seek to find the plane which $\Dkt$ is maximal and also passes through the origin of k-space, so we maximize $\Dkt$ with respect to angle $\phi$ about the $\kz$-axis. We need not consider the equivalent angle about the $\ky$-axis, as the tomogaphic CTF is symmetric.
\begin{equation*}
    \frac{\partial \Dkt}{\partial \phi} = 2 \kr \cos \phi \sin \frac{\Dt}{2}
\end{equation*}
Setting $\partial \Dkt / \partial \kx = 0$, we find an extremum when $\phi = \pi/2$ (along with trivial extrema $\kr=0$ and $\Dt = 0$). The second derivative test shows that $\phi = \pi/2$ is a maximum, meaning that $\Dkt$ is largest when $\kr$ is in the $\kx=0$ plane, making the problem 2D. The tomography CTF is formed by rotating individual CTFs about the $\kx$-axis, so $\krho$ (a polar coordinate representing distance from the tilt axis) can be substituted for $\kr$. The spacing between adjacent CTFs now simplifies to
\begin{equation}
    \Dkt =2 \krho \left ( \sin \frac{\Dt}{2} + \frac{\lambda}{2} \left ( \krho - \frac{2\alpha}{\lambda} \right ) \cos \frac{\Dt}{2} \right ) \label{eq:CTFdistanceNoApprox}
\end{equation}
The maximum measurable object size is inversely related to the distance between CTFs  as $D = 1/\Dkt$. We can use equation~\ref{eq:CTFdistanceNoApprox} to define the maximum size, $D$, of a reconstruction at a given convergence semi-angle, electron wavelength, tilt step, and maximum spatial frequency.
\begin{equation}
    D= \frac{1}{\lambda \krho^2 \left (\cos\frac{\Dt}{2} + \frac{1}{\lambda \krho} ( 2 \sin\frac{\Dt}{2} - 2 \alpha \cos\frac{\Dt}{2} )\right )} \label{eq:kFOVnoApprox}
\end{equation}
Introducing $d = 1/\krho$ to relate maximum spatial frequency to resolution, we change~\ref{eq:kFOVnoApprox} to 
\begin{equation}
    D = \frac{d^2}{\lambda \left ( \cos \frac{\Dt}{2} + \frac{d}{\lambda} (2 \sin \frac{\Dt}{2} - 2 \alpha \cos \frac{\Dt}{2}) \right )}\label{eq:dFOVnoApprox}
\end{equation}
Complete information transfer occurs when adjacent CTFs overlap. The distance between adjacent CTFs is zero at this point (labelled $\kc$ in Fig. \ref{fig:NewCrowther}), so we set Equation~\ref{eq:CTFdistanceNoApprox} to zero to find $\kc$.
\begin{equation*}
    \kc = \frac{2}{\lambda} \left ( \alpha - \tan \frac{\Dt}{2} \right )
\end{equation*}
Under the small angle approximation $\tan(\frac{\Dt}{2}) \approx \frac{\Dt}{2}$,
\begin{equation}
    \kc = \frac{2 \alpha - \Dt}{\lambda}
    \label{eq:kcEq}
\end{equation}
Equations~\ref{eq:kFOVnoApprox} and~\ref{eq:dFOVnoApprox} are only valid for $\kc \leq \krho \leq \km$ or $1/\km \leq d \leq 1/\kc$. For $0 \leq \krho \leq \kc$ and $1/\kc \leq d$, complete information is collected and the maximum object size is unbounded, giving piecewise equations. Using first-order small angle approximations ($\sin(\Dt/2) \approx \Dt/2$ and $\cos(\Dt/2) \approx 1$), equation~\ref{eq:kcEq}, and piecewise definition of maximum object size, equations~\ref{eq:kFOVnoApprox} and~\ref{eq:dFOVnoApprox} become, respectively,
\begin{gather}
    D = \left \{\begin{array}{ll}
        \infty, & 0 \leq \krho \leq \kc\\
        \frac{1}{\lambda \krho^2 (1 - \frac{\kc}{\krho})}, & \kc < \krho \leq \frac{2 \alpha}{\lambda}
        \end{array}
        \right . \label{eq:kNewCrowther}\\
    D = \left \{\begin{array}{ll}
        \infty, & \infty > d \geq \frac{1}{\kc}\\
        \frac{d^2}{\lambda (1 - k_c d)}, & \frac{1}{\kc} > d \geq \frac{\lambda}{\alpha}\\
        \end{array}
        \right . \label{eq:dNewCrowther}
\end{gather}
Equations $\ref{eq:kNewCrowther}$ and $\ref{eq:dNewCrowther}$ define the relationship between object size and resolution for aberration-corrected tomography analogous to the Crowther limit.
\section{Sampling From of Single Focal Plane}
\label{app:defocusSample}

\begin{figure}
    \centering
    \includegraphics[width = 244 pt]{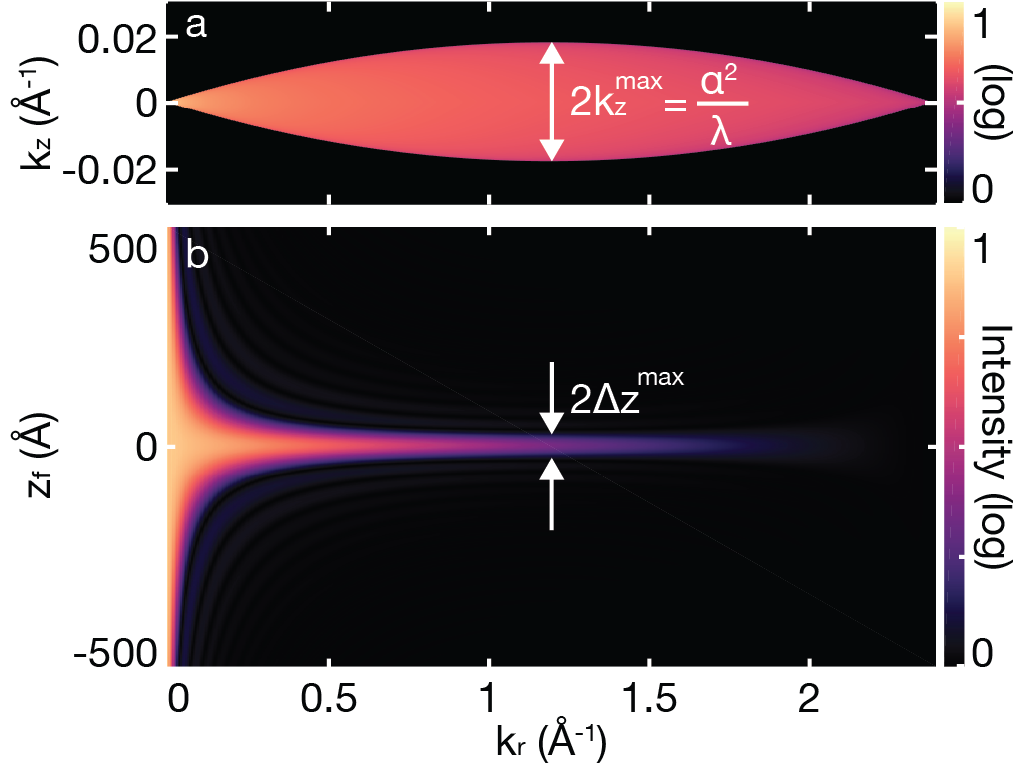}
    \caption{a) CTF of a through-focal image stack and the b) CTF Fourier transform along $k_z$ for a 200keV, 30mrad aberration-free electron beam. This illustrates that the midband frequencies set the maximum defocus step required for sampling with defocus. When the beam is out of focus from a specimen feature only low frequencies are transferred. Surprisingly, the highest frequencies are also transferred but the information intensity is too low to be useful.}
    \label{fig:zNyquist}
\end{figure}

To understand how information is sampled using defocus, consider a single projection taken at focal plane, $\Delta z$. Under an incoherent linear imaging model an image, $\mathcal{I}$, at defocus $\Delta z$ is a slice of the convolution of the object, $\mathcal{O} (\mathbf{r})$, with the electron probe, $h(\mathbf{r})$ :
\begin{alignat}{4}
    \mathcal{I} (\mathbf{r}) &= \left [ \mathcal{O} (\mathbf{r}) \right .&& \otimes && \left .h(\mathbf{r}) \right ] &&\cdot \delta(z - \Delta z ) \label{eq:projReal}\\
    &= \left [ \NPfig \right . && \otimes && \left .\PSFfig \right ] &&\cdot \planeFig \nonumber \\
    \mathcal{I} (\mathbf{k}) &= \left [ \mathcal{O} (\mathbf{k}) \right . && \ \cdot && \left . h (\mathbf{k}) \right ] && \otimes e^{-i \kz \Delta z} \delta(\kx) \delta(\ky) \label{eq:projK}\\
    &= \left [ \NPkfig \right . && \ \cdot && \left . \CTFfig \right ] && \otimes \rodFig \nonumber
\end{alignat}
In k-space, this is a multiplication followed by convolution with a rod whose phase oscillates according to defocus ( Eq.~\ref{eq:projK}. Evaluating the convolution in Equation~\ref{eq:projK} and rearranging, we find 
\begin{equation}
    \mathcal{I}(\kx, \ky, \Delta z) =\! \int_{- \infty} ^\infty d k'_z \ \mathcal{O} (\kx, \ky, k'_z) h (\kx, \ky, k'_z) e^{-i \Delta z k'_z}
\end{equation}
This is a Fourier transform in only one dimension from $k_z$-space to $\Delta z$-space. Figure~\ref{fig:zNyquist} shows this function for an aberration free beam and a point object---it illustrates how information in the midband is only measured within a limited focal range (2$\Delta z^{max}$). Nyquist sampling sets the depth resolution at $\Delta z^{max}=1/k_z^{max}= \lambda/\alpha^2$ which defines the depth-of-focus---derived herein using wave optics.
\section{Poissonian--Gaussian Noise Modeling of S/TEM Images}
\label{app:noisemodel}
An experimentally measured noisy S/TEM image, $\widetilde{p}(x,y)$, of a projected specimen, $p(x,y)$, is adequately modeled with both Poisson and Gaussian noise. With dose rate $\rchi$, acquisition time $t$, a noisy image becomes
\begin{gather}
    \label{eq:noiseModel}
    \widetilde{p}(x,y) = \rchi t \ p(x,y) + \rchi t \ n_p(0, \rchi t p(x,y)) + t n_g(0, \frac{\sigma^2}{t})
\end{gather}
The first two terms describe Poisson statistics and the last term follows Gaussian statistics \cite{foi_2008}. The Poisson noise is a function of both the specimen and dose ($\rchi t$) and the noise term $n_p$ is mean centered. The Gaussian noise $n_g$ is specimen independent, dose independent, mean-centered, and has variance $\frac{\sigma^2}{t}$ described by the central limit theorem. The measured image is the expected value of our measurement, $\E{\widetilde{p}(x,y)} = \rchi t \ p(x,y)$, and noise adds signal variance, $\Var{\widetilde{p}(x,y)} = \rchi t \ p(x,y) + t\sigma^2$. The variance has two terms from the gaussian and Poisson noise.

The signal to noise ratio (SNR) of a S/TEM image is:
\begin{align}
    \SNR{\widetilde{p}(x,y)} & = \frac{\E{\widetilde{p}^2(x,y)}}{\Var{\widetilde{p}(x,y)}} = 1 + \frac{ (\E{\widetilde{p}(x,y)})^2}{\Var{\widetilde{p}(x,y)}} \\ \nonumber
    & = 1+\frac{\rchi^2 t^2 \, p^2(x,y)}{\rchi t \, p(x,y) + t \sigma^2}
\end{align}
A S/TEM image is Poisson limited ($\rchi t \, p(x,y) \gg t \sigma^2$) for large signals or high detector efficiency and the SNR depends only on the object and total dose.
\begin{gather}
    \SNR{\widetilde{p}(x,y)} = 1 + \rchi t \, p(x,y)
\end{gather}

\section{Dose Fractionation for Through-Focal Imaging}
\label{app:doseFrac}
Here we show the SNR of useful information in a through-focal stack of images is only dependent on the total dose, so long as the through-focal stack is oversampled along defoci. Each image is taken at a beam defocus value ($z_f$) and, for a through-focal acquisition oversampled by a factor of $M$ ($M\, \Delta z < \lambda / \alpha^2$), $M$ adjacent defoci can be summed without any loss of information. $\delta z$ is the defocus step size. The SNR after summing $M$ adjacent defoci, each with dose-per-image $\rchi t$ at defoci $z_f+ m\, \delta z$, describes the quality of useful information. Because each acquired image is independently measured the expected value and variance of the sum adds linearly: $\E{\sum_{m} \widetilde{p}(x,y,z_f+m\, \Delta z)} = M \rchi t \, \bar{p}(x,y,z_f)$, $\Var{\sum_{m} \widetilde{p}(x,y,z_f+m\, \Delta z)} = M \rchi t \, \bar{p}(x,y,z_f) +Mt\sigma^2$. $\bar{p}(x,y,z_f)$ denotes the averaged image. Therefore, SNR after summing adjacent defocused images is 
\begin{align}
    \SNR{  \sum_{m} \widetilde{p}(x,y,z_f+m\, \Delta z)} = 1 + \frac{(M \rchi t\, \bar{p}(x,y,z_f) )^2}{M \rchi t\, \bar{p}(x,y,z_f) +Mt\sigma^2 }  
\end{align}

For Poisson noise limited images the SNR becomes
\begin{align}
    \SNR{  \sum_{m} \widetilde{p}(x,y,z_f+m\, \Delta z)} = 1 + M \rchi t\, \bar{p}(x,y,z_f)
\end{align}
Therefore, the SNR of oversampled through-focal stack only depends on the total dose ($M \rchi t$).

\section*{Supplementary Materials}
Appendix SI1-SI3\par
Figure SI1-SI4\par
Equation SI1-SI5

\pagebreak
\clearpage
\onecolumngrid
\renewcommand{\figurename}{SFIG.}
\renewcommand\thefigure{SI\arabic{figure}}
\renewcommand\thesection{SI\arabic{section}}
\renewcommand\theequation{SI\arabic{equation}}
\setcounter{section}{0}
\setcounter{figure}{0}
\setcounter{equation}{0}
\begin{center}{ {\Large Supplemental Information for \\
The Limits of Resolution and Dose for Aberration-Corrected Electron Tomography \\{~}\\
\normalsize Reed Yalisove, Suk Hyun Sung, Peter Ercius, and Robert Hovden}}

\end{center}

\section{3D Contrast Transfer Function of Highly Convergent Electron Beams}
\label{CTFderive}

Following the work from Intaraprasonk, Xin, and Muller \cite{Varat_Muller_2008}, we start with the point spread function (PSF). This is found by taking the inverse Fourier transform of a disk of radius $\km$ in k-space, then taking the magnitude and normalizing the result. 

The aberration and defocus of the beam can be accounted for by including an aberration function, $\chi(\kr, z)$ in the PSF. The PSF is found with an inverse Fourier transform of an aperture:
\begin{equation*}
    f(\kr) = \sqrt{\frac{1}{\pi \km^2}} \begin{cases}
    1, & 0 \leq \kr \leq \km \\
    0, & \textrm{otherwise}
    \end{cases}
\end{equation*}
The aperture is normalized such that $\int_{-\infty}^\infty|f(\kr)|^2 = 1$. Adding the aberration function and taking a Fourier transform, 
\begin{align*}
    \psi (r, z) &= \left | \frac{1}{(2\pi)^2} \int_0^{\km} \kr d\kr \int_0^{2\pi} d\theta \sqrt{\frac{1}{\pi \km^2}} e^{-i\chi(\kr, z)} e^{-i r \kr \cos(\theta)} \right |^2\\
    &= \frac{1}{\pi \km^2} \frac{1}{4 \pi^2} \left | \int _0 ^{\km} \kr d \kr J_0 (r \kr) e^{-i \chi (\kr, z)}\right | ^2
\end{align*}
The 3D contrast transfer function (CTF) of the electron beam is simply the Fourier transform of the PSF.
\begin{equation*}
    \Psi(\kr, \kz) = 2\pi \int _0 ^{\infty} r dr \int _{- \infty} ^{\infty} dz \psi (r, z) e^{i z \kz} J_0(r \kr)
\end{equation*}
To complete this integral, first expand the magnitude of the PSF using dummy variables $k_1$ and $k_2$. 
\begin{align*}
    \Psi(\kr,\kz) &= \frac{1}{2\pi^2\km^2} \int _0 ^{\infty} r dr \int _{- \infty} ^{\infty} dz  e^{i z \kz} J_0(r \kr)
    \left ( \int _0 ^{\km} k_1 d k_1 J_0 (r k_1) e^{-i \chi (k_1, z)}\right ) ^{\ast} 
    \left ( \int _0 ^{\km} k_2 d k_2 J_0 (r k_2) e^{-i \chi (k_2, z)} \right ) \\
    &= \frac{1}{2\pi^2\km^2} \int _0 ^{\infty} r dr \int _{- \infty} ^{\infty} dz e^{i z \kz} J_0(r \kr)
    \int _0 ^{\km} k_1 d k_1 \int _0 ^{\km} k_2 d k_2 J_0(r k_1) J_0 (r k_2)
    e^{i ( \chi (k_1,z) - \chi (k_2,z))}
\end{align*}
As noted by \cite{kirkland}, we can split the aberration function into a $z$-dependent and non-$z$-dependent component. The $z$-dependent term is defocus. Defining $z= 0$ as the in-focus plane, we get $\chi(k,z) = - \lambda z k^2/2 + C(k)$, where $C(k)$ represents the higher order aberrations. Substituting this into the previous equation, we find that 
\begin{align*}
    \Psi(\kr,\kz) &= \frac{1}{2\pi^2\km^2} \int _0 ^{\infty} r dr \int _{- \infty} ^{\infty} dz e^{i z \kz} J_0(r \kr)
    \int _0 ^{\km} k_1 d k_1 \int _0 ^{\km} k_2 d k_2 J_0(r k_1) J_0 (r k_2)
    e^{i \lambda z (k_2^2 - k_1^2)/2} e^{i(C(k_1) - C(k_2))}
\end{align*}
To complete this integral, we assume that all higher order aberrations are 0, i.e. $C(k) = 0$. First we integrate with respect to $z$.
\begin{align*}
    \int_{-\infty}^{\infty} dz e^{i z \kz} e^{i \lambda z (k_2^2 - k_1^2)/2} &= \delta \left ( \kz + \frac{\lambda}{2} (k_2^2 - k_1^2) \right ) \\
    &= \frac{1}{\lambda k_1} \left [ \delta \left(k_1 - \sqrt{\frac{2 \kz}{\lambda} + k_2^2} \right ) + \delta \left(k_1 + \sqrt{\frac{2 \kz}{\lambda} + k_2^2} \right ) \right ]
\end{align*}
This uses two identities: $\delta (ax) = \frac{1}{|a|} \delta (x)$ and $\delta (x^2 - a^2) = \frac{1}{2a} ( \delta (x - a) + \delta (x + a))$.
We can then integrate with respect to $k_1$. 
\begin{align*}
    &\int _0 ^{\km} k_1 dk_1 J_0(r k_1) \frac{1}{\lambda k_1} \left [ \delta \left(k_1 - \sqrt{\frac{2 \kz}{\lambda} + k_2^2} \right )  + \delta \left(k_1 + \sqrt{\frac{2\kz}{\lambda} + k_2^2} \right ) \right ]
\end{align*}
The second term in this integral is always $0$, and the first term is only nonzero when $\km \leq \sqrt{\frac{2 \kz}{\lambda} + k_2^2}$, which implies $k_2 \leq \sqrt{\km^2 - \frac{2 \kz}{\lambda}} (\ast)$. The value of the integral is
\begin{align*}
    &\frac{1}{\lambda}\int_0^{\km} dk_1 J_0(r k_1) \delta \left(k_1 - \sqrt{\frac{2 \kz}{\lambda} + k_2^2} \right ) = \frac{1}{\lambda} J_0 \left ( r\sqrt{\frac{2 \kz}{\lambda} + k_2^2} \right ) 
\end{align*}

We will next integrate with respect to $r$. 

\begin{align*}
    \Psi (\kr, \kz) &= \frac{1}{2\pi^2 \lambda \km^2} \int _0 ^{\infty} r dr \int _0 ^{\km} k_2 dk_2 \nonumber
    \times J_0(r k_r) J_0(r k_2) J_0 \left ( r\sqrt{\frac{2 \kz}{\lambda} + k_2^2} \right )
\end{align*}

From \cite{zwillinger}, we find the formula

\begin{align*}
    \int _0 ^{\infty} x dx J_0 (ax) J_0 (bx) J_0 (cx) = \frac{1}{2 \Delta \pi}
\end{align*}

Here, $\Delta$ is the area of a triangle with sidelengths $a$, $b$, and $c$, given by $\Delta = \frac{1}{4} \sqrt{[c^2 - (a - b)^2][(a + b)^2 - c^2]}$. Here, $a = \kr$, $b = k_2$, and $c = \sqrt{2 \kz / \lambda + k_2 ^2}$. This allows us to show
\begin{align}
    \Psi(\kr, \kz) &= \frac{1}{\pi^3 \lambda \km ^2 \kr} \int _0 ^{\km} k_2 dk_2 \times \left ( 4 k_2 ^ 2 - \left ( \frac{2 \kz}{\lambda \kr} - \kr \right ) ^2 \right ) ^{-1/2} \label{k2int}
\end{align}
This integral is nonzero when the triangle inequality holds for $a$, $b$, and $c$. In particular, we are interested in two of the three possible inequalities:
\begin{align*}
    \kr + k_2 & \geq \sqrt{\frac{2 \kz}{\lambda} + k_2 ^2} \\
    k_2 + \sqrt{\frac{2 \kz}{\lambda} + k_2 ^2} & \geq \kr 
\end{align*}
These can be rearranged to show that
\begin{equation}
    k_2 \geq \left | \frac{\kz}{\kr \lambda} - \frac{\kr}{2} \right | \label{k2ineq}
\end{equation}
From $(\ast)$ and Eq.~\ref{k2ineq} we see that the bounds on the integral in Eq.~\ref{k2int} reduce to 
\begin{align}
    \Psi(\kr, \kz) &= \frac{1}{\pi^3 \lambda \km ^2 \kr} \int _{\frac{\kz}{\kr \lambda} - \frac{\kr}{2}} ^{\sqrt{\km^2 - \frac{2 \kz}{\lambda}}} k_2 dk_2
     \times \left ( 4 k_2 ^ 2 - \left ( \frac{2 \kz}{\lambda \kr} - \kr \right ) ^2 \right ) ^{-1/2} \nonumber\\
    &= \left . \frac{1}{4 \pi^3 \lambda \km ^2 \kr} \sqrt{ 4 k_2 ^2 - \left (\frac{2 \kz}{\lambda \kr} - \kr \right ) ^2} \ \right | _{\frac{\kz}{\kr \lambda} - \frac{\kr}{2}} ^{\sqrt{\km^2 - \frac{2 \kz}{\lambda}}} \nonumber\\
    &= \frac{1}{2 \pi^3 \lambda \km \kr} \sqrt{ 1 - \left ( \frac{\kr}{2 \km} + \frac{\kz}{\lambda \km \kr}\right )^2} \label{CTFkm}
\end{align}
Using the identity $\alpha = \lambda \km$ from \cite{kirkland}, where $\alpha$ is the aperture semi-angle, Eq.~\ref{CTFkm} becomes 
\begin{equation}
    \Psi(\kr, \kz) = \frac{1}{2 \pi^3 \alpha \kr} \sqrt{ 1 - \left ( \frac{\lambda \kr}{2 \alpha} + \frac{\kz}{\alpha \kr}\right )^2} \label{CTF}
\end{equation}
The leading coefficient of this result varies slightly from \cite{Varat_Muller_2008} due to differences in the Fourier transform convention. We can find the bounds on $\kr$ and $\kz$ for which this equation is nonzero by consider Eq.~\ref{k2ineq} and $(\ast)$. Combining these equations, we find that 
\begin{align}
    \frac{2 \kz}{\lambda} & \leq 2 \kr k_2 + \kr^2 \leq 2 \kr \sqrt{\km ^2 - \frac{2 \kz}{\lambda}} + \kr ^2 \nonumber \\
    \left (\frac{2 \kz}{\lambda} - \kr ^2 \right ) ^2 &\leq \left ( 2 \kr \sqrt{\km ^2 - \frac{2 \kz}{\lambda}} \right ) ^2 \nonumber \\
    \left (\frac{2 \kz}{\lambda} + \kr ^2 \right ) ^2 &\leq 4 \kr ^2 \km ^2 \nonumber \\
    |\kz| &\leq - \frac{ \lambda}{2} \kr \left (\kr -  2 \km \right ) \nonumber \\
    |\kz| &\leq - \frac{\lambda}{2} \kr \left (\kr - \frac{2 \alpha}{\lambda} \right ) \label{CTFbound}
\end{align}

\section{Methods: Tomographic Simulation using Quantum Mechanical Multislice Scattering}
\label{SuppMethods}
\begin{figure}[ht]
    \centering
    \includegraphics[width = 468pt]{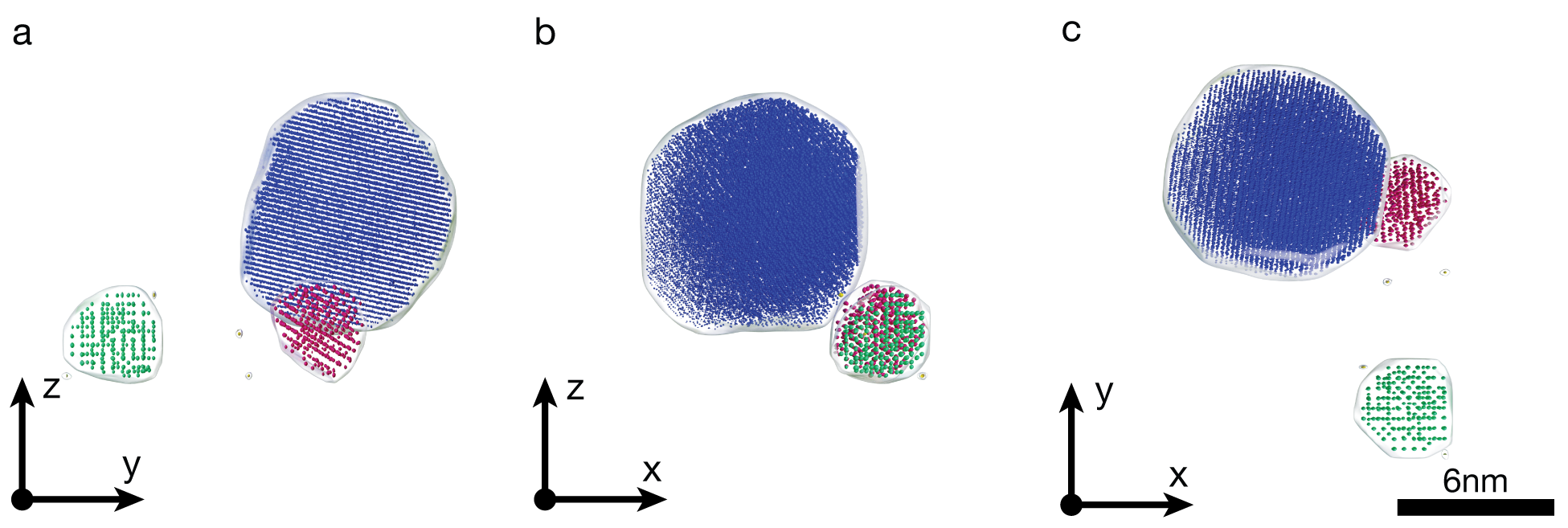}
    \caption{3D atomic resolution aberration corrected tomographic reconstruction of from quantum mechanical multislice simulation. a,b,c) Orthographic view of the 3D reconstruction highlights atomic resolution in multiple viewing angles.}
    \label{fig:recon}
\end{figure}
Fully quantum mechanical multislice simulation of three synthetic FePt nanoparticles spanning (15nm)$^3$ was performed at incident electron energy of 200keV and convergence semi-angle of 30mrad, pixel size (0.25\AA)$^2$ using PRISMATIC software \cite{Pryor_2017}. Images were calculated on GPU accelerated computing clusters at University of Michigan Advanced Research Computing and Technology Center and Lawrence Berkeley National Labs. The atomic coordinates for the FePt nanoparticles used were experimentally acquired by Yang et al. \cite{FePt_MDB}. Each through-focal stack contains 13 defoci images with a 1.25nm defocus step; 105 through-focal stacks were simulated at each tilts with a 30mrad ($1.714\degree$) tilt step. In creating over 1300 images, the simulation computed ~500 million wavefunctions over 15,000 GPU core hours.

To weight the through-focal stack in Fourier space, it is divided by the aberration-free CTF at 300keV, 30mrad, this is multiplied with distance from the tilt axis (analogous to weighted back projection). The weighted through-focal stack from each tilt angle is mapped onto a universal Fourier space by bilinear extrapolation, which distributes the complex value of an input point to its four nearest neighbors on the output Cartesian grid. The final reconstruction is obtained directly from the 3D inverse Fourier transform.

\section{Additional figures}
\label{addFigs}
\begin{figure}[ht]
    \centering
    \includegraphics[width = 244pt]{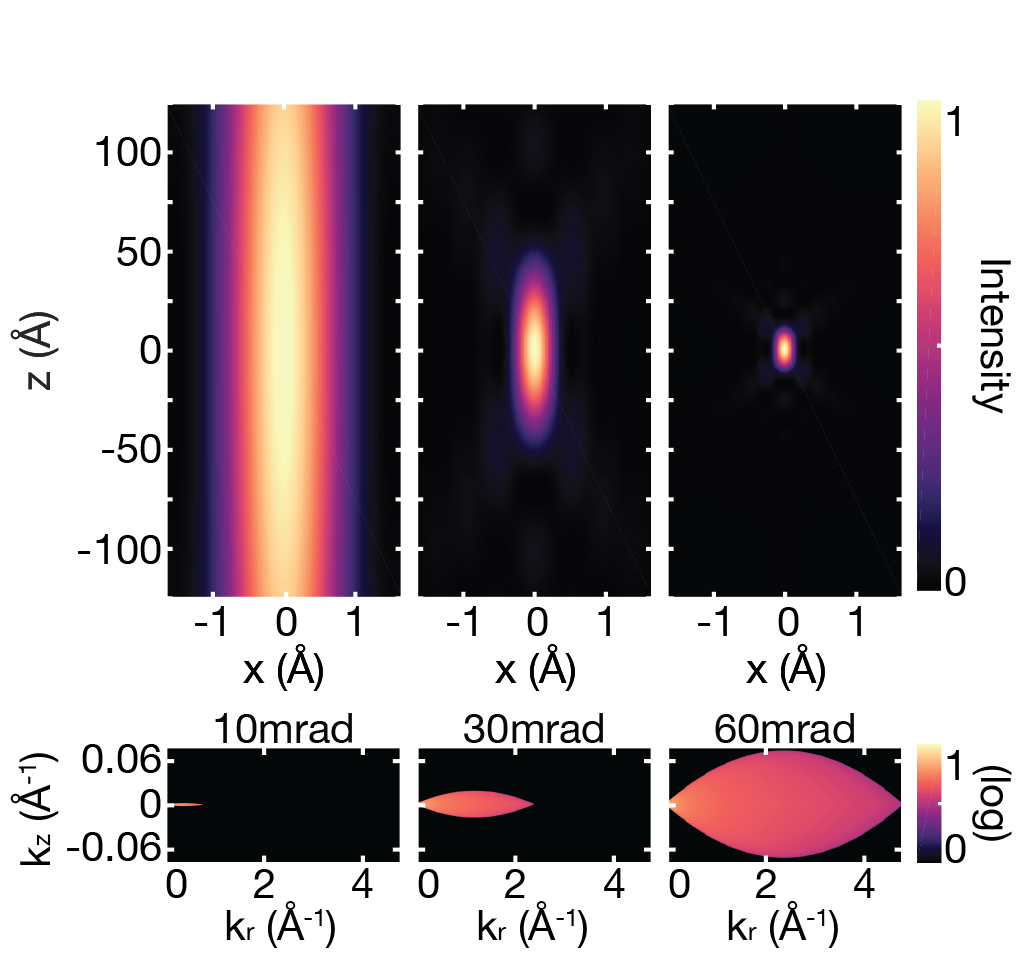}
    \caption{2D slices through the PSF (top) and CTF (bottom) of electron wave functions with 10, 30, and 60mrad aperture semiangles. Confinement of the PSF in the z-direction corresponds to extent in the $\kz$-direction of the CTF.}
    \label{fig:psfctf}
\end{figure}

\begin{figure*}[ht]
    \centering
    \includegraphics[width = 419pt]{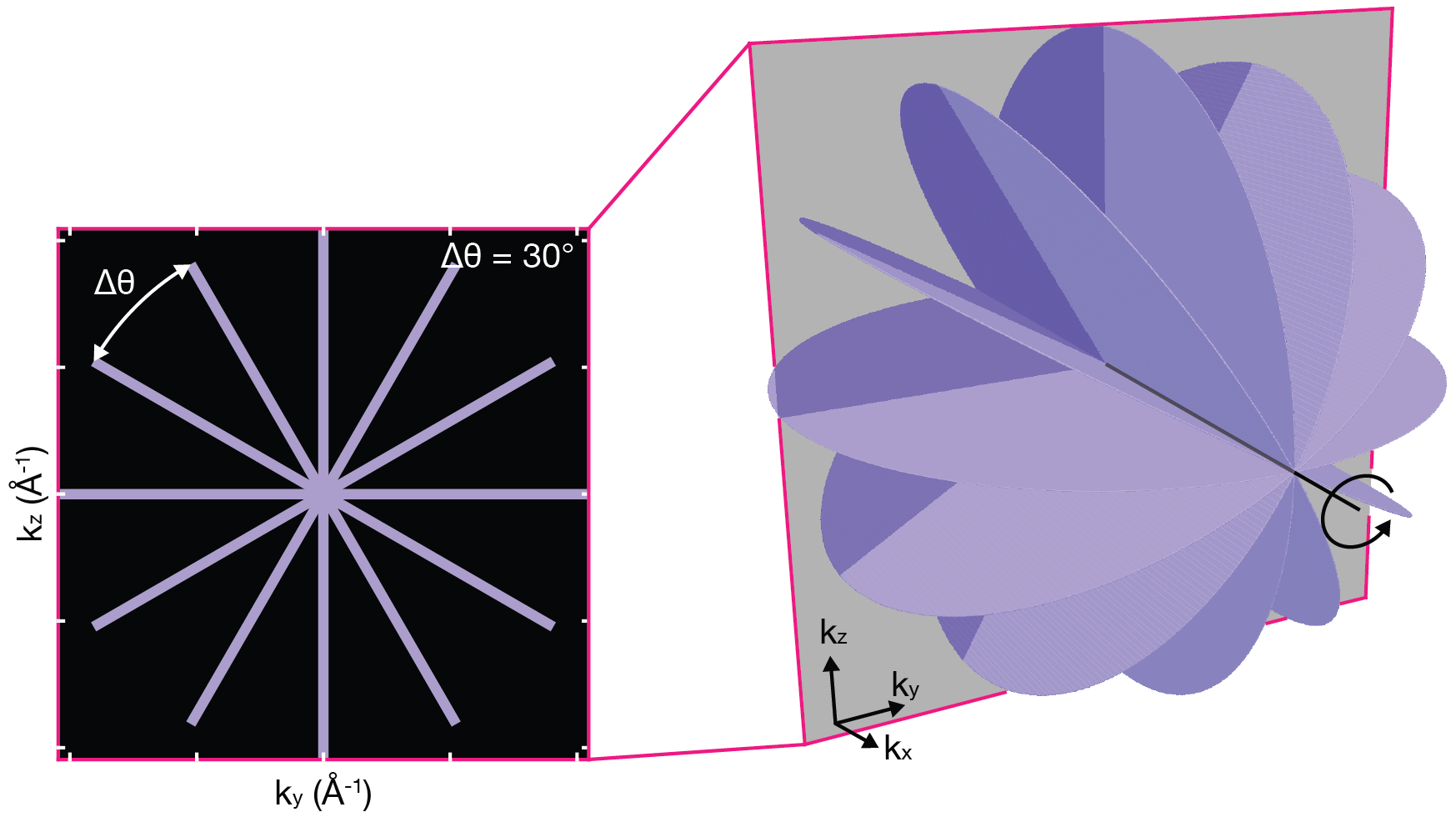}
    \caption{3D Contrast Transfer Function of the CTF for conventional tomography. The left image is a slice through the $\ky \kz$-plane of the 3D CTF.}
    \label{fig:convCTF3D}
\end{figure*}

\begin{figure*}[ht]
    \centering
    \includegraphics[width = 237pt]{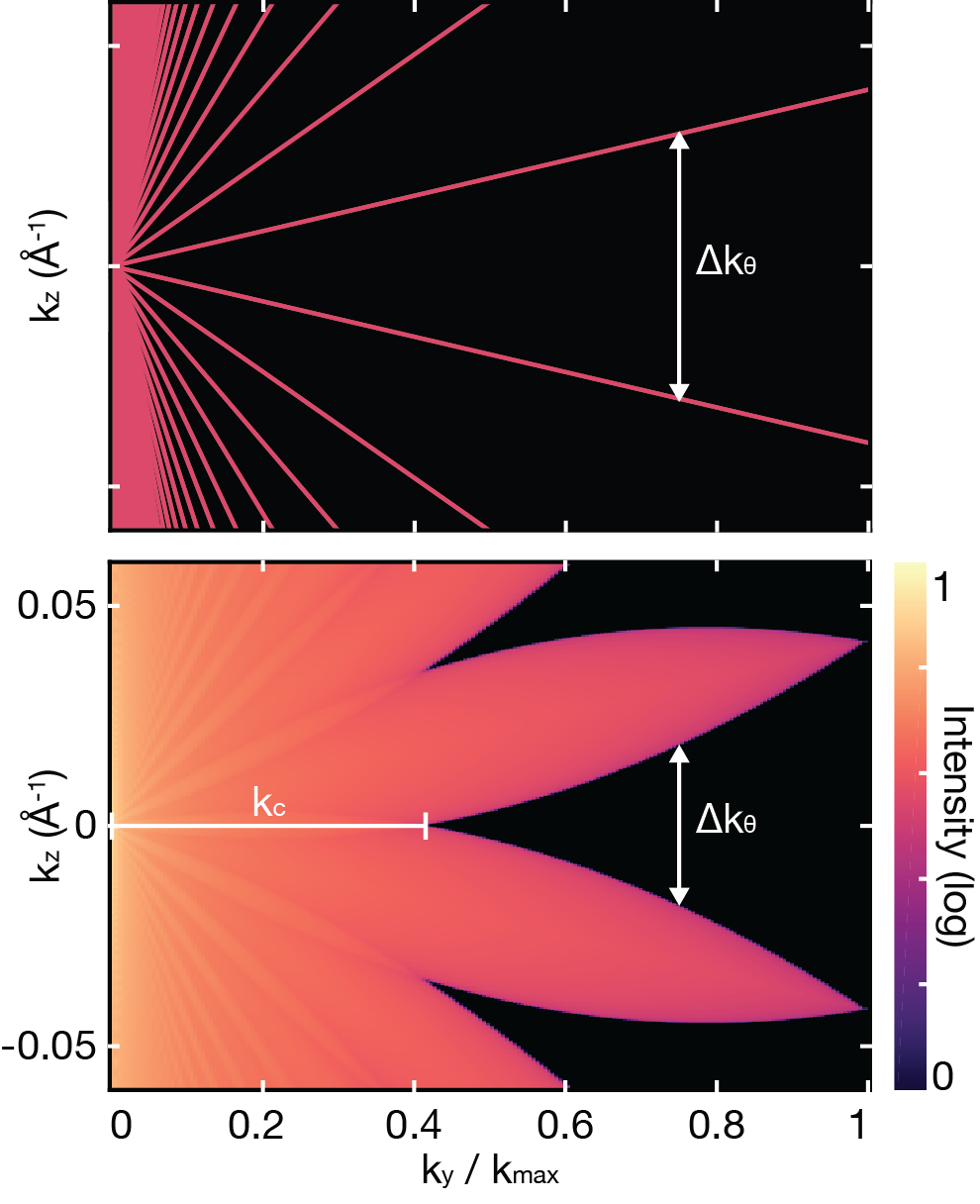}
    \caption{Tomography CTFs for a) traditional tomography with $2^\circ$ tilt, where information is collected along planes and b) aberration-corrected tomography with 200keV, $\alpha= 30$mrad, and $2^\circ$ tilts. In aberration corrected tomography, information is completely sampled for frequencies below $\kc$. In each figure, the distance between adjacent regions of information ($\Dkt$) is labelled. This distance is larger for traditional tomography, leading to smaller allowed object size}
    \label{fig:convVsabCorr}
\end{figure*}

\end{document}